\documentclass[twocolumn,preprintnumbers,amsmath,amssymb]{revtex4-2}
\usepackage{color}
\usepackage{bm, graphicx, amsmath}
\usepackage{hyperref}

\usepackage{algorithm}
\usepackage{algpseudocode}
\usepackage{fancyhdr}
\begin{document}
	
	\fancyhead[R]{\ifnum\value{page}<2\relax\else\thepage\fi}
	\title{Variational Amplitude Amplification for Solving QUBO Problems}
	\author{Daniel Koch$^{1}$$^{*}$, Massimiliano Cutugno$^{1}$, Saahil Patel$^{1}$, Laura Wessing$^{1}$, Paul M. Alsing$^{1}$ }
	\affiliation{$^{1}$Air Force Research Lab, Information Directorate, Rome, NY }
	\affiliation{$^{*}$Corresponding Author: daniel.koch.13@us.af.mil  }
	
	\begin{abstract}
		
		We investigate the use of amplitude amplification on the gate-based model of quantum computing as a means for solving combinatorial optimization problems.  This study focuses primarily on QUBO (quadratic unconstrained binary optimization) problems, which are well-suited for qubit superposition states.  Specifically, we demonstrate circuit designs which encode QUBOs as `cost oracle' operations $U_{\textrm{C}}$, which when combined with the standard Grover diffusion operator $U_{\textrm{s}}$ lead to high probabilities of measurement for states corresponding to the optimal and near optimal solutions.  In order to achieve these probabilities, a single scalar parameter $p_{\textrm{s}}$ is required, which we show can be found through a variational quantum-classical hybrid approach.
		
	\end{abstract}

	\maketitle
	
	\thispagestyle{fancy}
	
	\section{Introduction}                                  
	
	Amplitude amplification is a quantum algorithm strategy that is capable of circumventing one of quantum computing's most difficult challenges: probabilistic measurements.  Originally proposed by Grover in 1996 \cite{grover}, and later shown to be optimal \cite{boyer,bennett}, the combination of his oracle $U_{\textrm{G}}$ and `diffusion' $U_{\textrm{s}}$ operators is able to drive a quantum system to a superposition state where one (or multiple) basis state(s) has nearly 100\% probability of being measured.  Since then, many researchers have contributed to the study of $U_{\textrm{G}}$ and $U_{\textrm{s}}$  \cite{farhi,brassard1,brassard2,childs,ambainis,singleton}, seeking to better understand how the fundamental nature of amplitude amplification is dependent on these two operators.  Similarly, the aim of this study is to further extend the capabilities of amplitude amplification as a means for solving combinatorial optimization problems using gate-based quantum computers.
	
	The results of this paper are a continuation of our previous work \cite{koch2}, in which we demonstrated an oracle design which was capable of encoding and solving a weighted directed graph problem.  The motivation for this oracle was to address a common criticism of $U_{\textrm{G}}$ \cite{lloyd,viamontes,reg,seidel,nielsen}, namely that the circuit construction of oracles too often hardcodes the solution it aims to find, negating the use of quantum entirely.  Similar to other recent studies \cite{bang,satoh,bench,shyamsundar,gilliam,roy,plek}, we showed that this problem can be solved at the circuit depth level by avoiding gates such as control-Z for constructing the oracle, and instead using phase and control-phase gates (P($\theta$) and CP($\theta$)).  However, simply changing the phase produced from $U_{\textrm{G}}$ to something other than $\pi$ is not enough \cite{long1,long2,hoyer,younes,li1,guo}.  Our oracle construction applies phases to not only a desired marked state(s), but $all$ states in the full $2^N$ Hilbert Space.  The phase each basis state receives is proportional to the solutions of a weighted combinatorial optimization problem, for which the diffusion operator $U_{\textrm{s}}$ can be used to boost the probability of measuring states that correspond to optimal solutions.
	
	The consequence of using an oracle operation that applies phases to every basis state is an interesting double-edged sword.  As we show in sections II. - IV., and later in section VII., the use of phase gates allows for amplitude amplification to encode a broad scope of combinatorial optimization problems into oracles, which we call `cost oracles' $U_{\textrm{c}}$.  In particular, we demonstrate the robustness of amplitude amplification for solving these kinds of optimization problems with asymmetry and randomness \cite{song,pomeransky,janmark}.  However, the tradeoff for solving more complex problems is twofold.  Firstly, in contrast to Grover's oracle, using $U_{\textrm{c}}$ is only able to achieve peak measurement probabilities up to $70$-$90$\%.  In section VI. we show that these probabilities are still high enough for quantum to reliably find optimal solutions, which notably are achieved using the same O( $\frac{\pi}{4} \sqrt{N/M}$ ) iterations as standard Grover's \cite{grover,boyer,bennett}.
	
	The second, more challenging tradeoff when using $U_{\textrm{c}}$ is that the success of amplitude amplification is largely dependant on the correct choice of a single free parameter  $p_{\textrm{s}}$ \cite{koch2}.  This scalar parameter is multiplied into every phase gate for the construction of $U_{\textrm{c}}$ (P($\theta \cdot p_{\textrm{s}}$) and CP($\theta \cdot p_{\textrm{s}}$)), and is responsible for transforming the numeric scale of a given optimization problem to values which form a range of approximately $2 \pi$.  This in turn is what allows for reflections about the average amplitude via $U_{\textrm{s}}$ to iteratively drive the probability of desired solution states up  to $70$-$90$\%.  The significance of $p_{\textrm{s}}$, and the challenges in determining it experimentally, are a major motivation for this study.  In particular, the results of section V. demonstrate that there is a range of $p_{\textrm{s}}$ values for which many optimal solutions can be made to become highly probable.  Additionally, our simulations show that there is an observed correlation between the numerical cost function value of these solutions and the $p_{\textrm{s}}$ values where they achieve peak probabilities.  This underlying correlation supports the idea of using amplitude amplification for a variational model of hybrid quantum-classical computing, which is the core finding of this study.

	\subsection{Layout}                                              
	
	The layout of this study is as follows.  Section II. begins with the mathematical formalism for the optimization problem we will seek to solve using amplitude amplification.  Sections III. \& IV. discuss the construction of the problem as a quantum circuit, the varying degrees of success one can expect from optimization problems generated using random numbers, and the conditions for which these successes can be experimentally realized.  In section V. we explore the role of $p_{\textrm{s}}$ from a heuristic perspective, whereby we demonstrate that many near optimal solutions are capable of reaching significant probabilities of measurement.  Section VI. is a primarily speculative discussion, theorizing how the collective results of section V. can be coalesced into a hybrid quantum-classical variational algorithm.  And finally, section VII. completes the study with additional optimization problems that can be constructed as oracles and solved using amplitude amplification. 
	
	\section{QUBO Definitions}%
	
	We begin by outlining the optimization problem which will serve as the focus for this study: QUBO (quadratic unconstrained binary optimization).  The QUBO problem has many connections to important fields of computer science \cite{kochenberger,lucas,glover,date,herman}, making it relevant for demonstrating quantum's potential for obtaining solutions.  To date, the two most successful quantum approaches to solving QUBOs are annealing \cite{date2,ushijima,pastorello,cruz} and QAOA \cite{qaoa,qaoa2,guerreschi,guerreschi2}, with a lot of interest in comparing the two \cite{streif,gabor,pelofske}.  Shown below in equation \ref{Eqn.1} is the QUBO cost function C(X) which we shall seek to solve using our quantum algorithm.
	
	\begin{eqnarray}            
		\textrm{C}(\textrm{X}) = \sum_i^N W_i x_i + \sum_{ \{i,j\} \hspace{0.02cm} \in \hspace{0.02cm} \mathbb{S}} w_{ij} x_i x_j
		\label{Eqn.1}
	\end{eqnarray} 
	
	The function C(X) evaluates a given binary string X of length $N$, composed of individual binary variables $x_i$.  Together, the total number of unique solutions to each QUBO is $2^N$, which is also the number of quantum states producible from $N$ qubits.  Throughout this study we will use subscripts X$_i$ and C(X$_{i}$) when referring to individual solutions, and C(X) when discussing a cost function more generally.
	
\begin{figure}[h]            
		\centering
		\includegraphics[scale=.6]{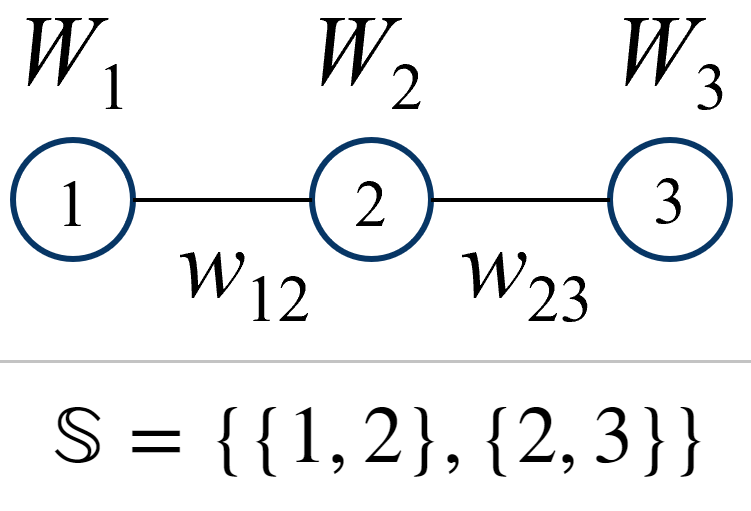}
		\caption{ (top) An example $3$-qubit linear QUBO with weighted nodes and edges. (bottom) The set $\mathbb{S}$ containing the complete connectivity of the QUBO.}
		\label{Fig.1}
	\end{figure}

	As shown in equation \ref{Eqn.1}, a QUBO is defined by two separate summations of weighted values.  The first summation evaluates weights $W_i$ associated with each individual binary variable, while the second summation accounts for pairs of variables which share a weighted connection $w_{ij}$.  In this study we adopt the typical interpretation of QUBOs as graph problems, whereby each binary variable $x_i$ represents a node.  We can then define the connectivity of a QUBO graph using the set $\mathbb{S}$, which itself is a collection of sets that describe each pair of nodes $x_i$ and $x_j$ that share a connection. See figure \ref{Fig.1} below for an example.

	The interest of this study is to use a quantum algorithm to find either X$_{\textrm{min}}$ or X$_{\textrm{max}}$, which are the solutions which minimize / maximize the cost function C(X) respectively.  For all QUBOs analyzed in the coming sections, the weight values $W_i$ and $w_{ij}$ are restricted to integers, randomly selected from a uniform distribution as shown below in equations \ref{Eqn.2} and \ref{Eqn.3}.
	
	\begin{eqnarray}            
		W_i , w_{ij} &\in& \mathbb{Z}  \label{Eqn.2} \\ 
		W_i , w_{ij} &\in& [-100 , 100]  \label{Eqn.3}
	\end{eqnarray} 
	
	In section V. we discuss the consequences of choosing weight values in this manner and its advantage for quantum.  However, nearly all of the results shown throughout this study are applicable to the continuous cases for $W_i$ and  $w_{ij}$ as well, with the one exception being the results of section V.D.

	\subsection{Linear QUBO}                                              
	
	The cost function given in equation \ref{Eqn.1} is applicable to any graph structure $\mathbb{S}$, so long as every node and edge is assigned a weight.  For this study we will focus on one specific $\mathbb{S}$, which we refer to as a `linear QUBO.'  The connectivity of these graphs is as follows:
	
	\begin{eqnarray}            
		\mathbb{S}  &=& \{   \{ n,n+1 \} \hspace{0.1cm} | \hspace{0.1cm} 1 \leq n \leq N-1  \}  \label{Eqn.4} 
	\end{eqnarray} 
	
	As the name suggests, linear QUBOs are graphs for which every node has connectivity with exactly two neighboring nodes, except for the first and final nodes.  The motivation for studying QUBOs of this nature is their efficient realizability as quantum circuits, given in the next section.
	
	\section{Amplitude Amplification}%
	
	The quantum strategy for finding optimal solutions to C(X) investigated in this study is amplitude amplification \cite{farhi,brassard1,brassard2,childs,ambainis,singleton}, which is the generalization of Grover's algorithm \cite{grover}.  The full algorithm is shown below in Alg. \ref{Alg.1}, which notably is almost identical to Grover's algorithm except for the replacement of Grover's oracle $U_{\textrm{G}}$ with our cost oracle $U_{\textrm{c}}$ .
	
	\begin{algorithm}[H]
	\caption{Amplitude Amplification Algorithm} \label{Alg.1}
	\begin{algorithmic}
		\State Initialize  Qubits: $|\Psi\rangle = |0\rangle ^{\otimes N}$
		\State Prepare Equal Superposition: $H^{\otimes N} |\Psi \rangle = |s\rangle$
		\For{ $k \approx  \frac{\pi}{4} \sqrt{ 2^N }$  }  
		\State Apply $U_\textrm{c} |\Psi\rangle$ (Cost Oracle)
		\State Apply $U_\textrm{s} |\Psi\rangle$ (Diffusion)
		\EndFor
		\State Measure
	\end{algorithmic}
	\end{algorithm}

	By interchanging different oracle operations into the Alg. \ref{Alg.1}, various problem types can be solved using ampltitude amplification.  For example, Grover's original oracle solves an unstructured search, whereas here we are interested in optimal solutions to a cost function.  Later in section VII. we discuss further oracle adaptations and the problems they solve.  For all oracles, we use the standard diffusion operator $U_{\textrm{s}}$, given below in equation \ref{Eqn.5}.
	
	\begin{eqnarray}            
		U_{\textrm{s}}  = 2 | s \rangle \langle s | - \mathbb{I}
		\label{Eqn.5}
	\end{eqnarray} 
	
	This operation achieves a reflection about the average amplitude, whereby every basis state in $| \Psi \rangle$ is reflected around their collective mean in the complex plane.  This operation causes states' distance from the origin to increase or decrease based on their location relative to the mean, which in turn determines their probability of measurement.  Therefore, a successful amplitude amplification is able to drive the desired basis state(s) as far from the origin as possible, up to a maximum distance of $1$ (measurement probability of 100$\%$).
	
	\subsection{Solution Space Distribution}%
	
	A prerequisite for the success of amplitude amplification as demonstrated in this study is an optimization problem's underlying solution space distribution; that is, the manner in which all possible solutions to the problem are distributed with respect to one another.  For QUBOs, these are the $2^N$ possible C(X$_i$) cost function values.  Shown below in figure \ref{Fig.2} is a histogram of one such solution space distribution, for the case of a length $20$ linear QUBO according to equations \ref{Eqn.1} - \ref{Eqn.4}. The x-axis represents all possible cost function evaluations, and the y-axis is the corresponding number of unique X$_i$ solutions that result in the same C(X$_i$) value.
	
	\begin{figure}[h]            
		\centering
		\includegraphics[scale=.38]{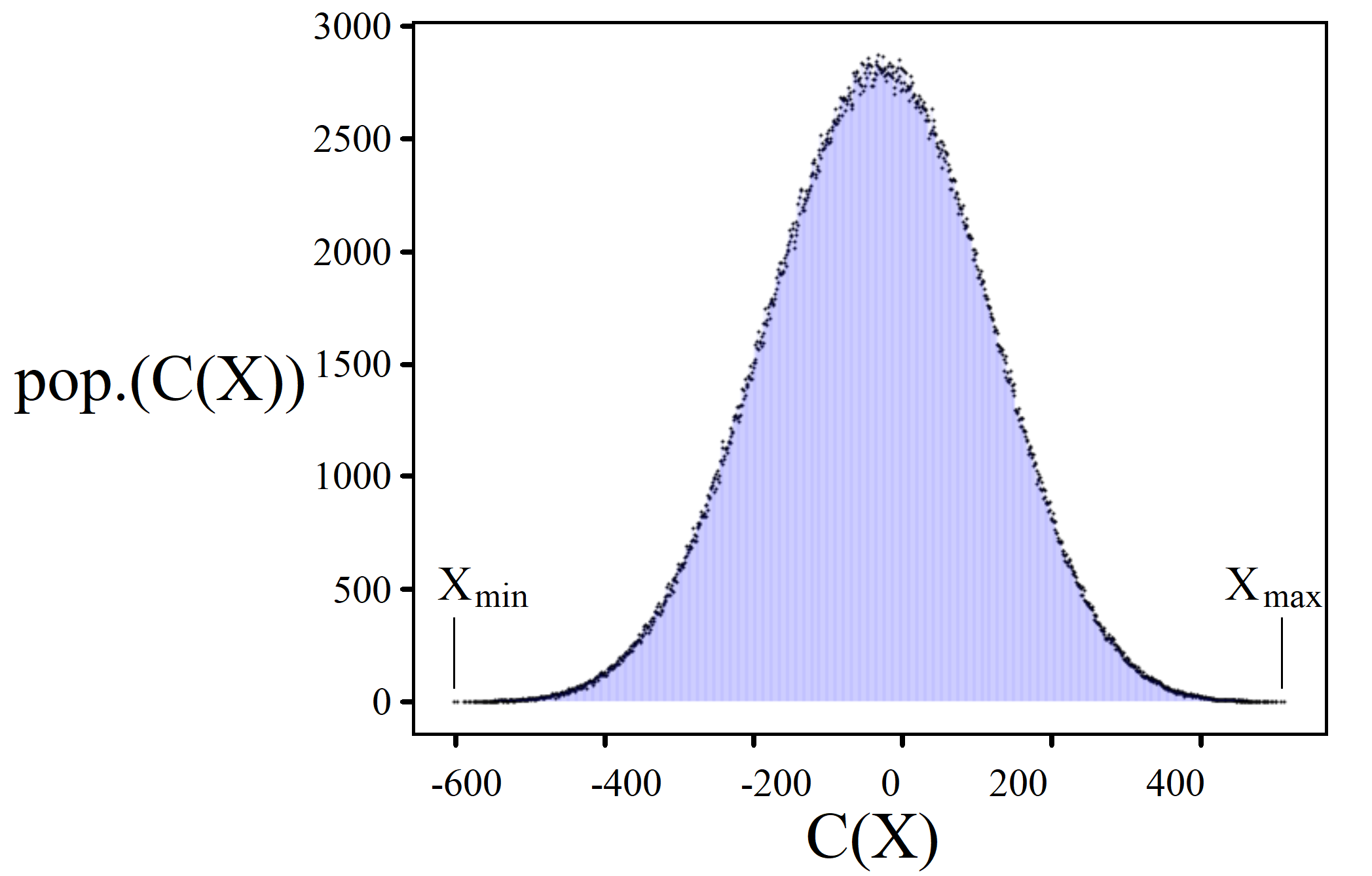}
		\caption{Example of a solution space distribution for a $20$ node linear QUBO, with weights according to equations \ref{Eqn.2} and \ref{Eqn.3}.  }
		\label{Fig.2}
	\end{figure}
	
	Depicted in figure \ref{Fig.2} are all $2^{20}$ possible solutions to an example linear QUBO.  Because this QUBO was generated from randomized weights, the combination of the Law of Large Numbers \cite{bernoulli} and Central Limit Theorem \cite{laplace} predicts that its underlying solution space should be approximately gaussian \cite{gauss} in shape, given by equation \ref{Eqn.6}.
	
	\begin{eqnarray}            
		\textrm{G}(x) = \alpha \hspace{0.02cm} \textrm{e}^{\frac{(x - \mu)^2}{2 \sigma ^2}}
		\label{Eqn.6}
	\end{eqnarray} 
	
	Indeed, the histogram shown is approximately gaussian, but importantly it has imperfections resulting from the randomized weights.  At large enough problem sizes (around $N \geq 20$), these imperfections have minimal impact on a problem's aptitude for amplitude amplification, which was a result from our previous study \cite{koch2}.   Similarly, another recent study \cite{bench} demonstrated that in addition to symmetric gaussians, solution space distributions for both skewed gaussians and exponential profiles also lead to successful amplitude amplifications.  The commonality between these three distribution shapes is that they all possess large clusters of solutions that are sufficiently distanced from the optimal solutions we seek to boost.  This can be seen in figure \ref{Fig.2} as the location of X$_{\textrm{min}}$ and X$_{\textrm{max}}$ as compared to the central peak of the gaussian.  When appropriately encoded as an oracle $U_{\textrm{c}}$, these clusters serve to create a mean point in the complex plane which the optimal solution(s) use to reflect about and increase in probability.

	\subsection{Cost Oracle $U_{\textrm{c}}$}%
	
	In order to use algorithm \ref{Alg.1} for finding the optimal solution to a given cost function, we must construct a cost oracle $U_{\textrm{c}}$ which encodes the weighted information and connectivity of the problem.  In our previous study we referred to this operation as a `phase oracle' $U_{\textrm{P}}$ \cite{koch2}, and similarly it has also been called a `subdivided phase oracle'  SPO \cite{satoh,bench} or `non-boolean oracle' \cite{shyamsundar}.  How one constructs $U_{\textrm{c}}$ is problem specific, but the general strategy is to primarily use two quantum gates, shown below in equations \ref{Eqn.7} and \ref{Eqn.8}.

	\begin{eqnarray}            
		\textrm{P}(\theta) &=& \begin{bmatrix} 1 & 0 \\ 0 & \textrm{e}^{i \theta} \end{bmatrix} \label{Eqn.7} \\
		\textrm{CP}(\theta) &=& \begin{bmatrix} 1 & 0 & 0 & 0 \\ 
			0 & 1 & 0 & 0 \\ 0 & 0 & 1 & 0 \\ 0 & 0 & 0 &  \textrm{e}^{i \theta} \end{bmatrix}\label{Eqn.8}
	\end{eqnarray} 
	
	The single and two-qubit gates P($\theta$) and CP($\theta$) are referred to as phase gates, also known as R$_{\textrm{z}}$($\theta$) and CR$_{\textrm{z}}$($\theta$) for their effect of rotating a qubit's state around the z-axis of the Bloch sphere.  Mathematically they are capable of applying complex phases as shown below.
	
	\begin{eqnarray}            
		\textrm{P}(\theta)  | 1 \rangle  &=& \textrm{e}^{i \theta} | 1 \rangle \label{Eqn.9} \\
		\textrm{CP}(\theta)  |  1 1 \rangle  &=& \textrm{e}^{i \theta} |  1 1 \rangle\label{Eqn.10}
	\end{eqnarray} 
	
	Applying P($\theta$) to a qubit only affects the $| 1 \rangle$ state, leaving $| 0 \rangle$ unchanged, and similarly only $| 11 \rangle $ for CP($\theta$).  However, this is exactly what we need in order to construct C(X) from equation \ref{Eqn.1}.  When evaluating a particular binary string X$_{i}$ classically, only instances where the binary values $x_i$ are equal to $1$ yield non-zero terms in the summations.  For quantum, each binary string X$_i$ is represented by one of the $2^N$ basis states $| \textrm{X}_i \rangle$.  Thus, our quantum cost oracle $U_{\textrm{c}}$ can replicate C(X) by using P($\theta$) and CP($\theta$) to only effect basis states with qubits in the $| 1 \rangle$ and $| 11 \rangle$ states.
	
	\begin{figure}[h]            
		\centering
		\includegraphics[scale=.6]{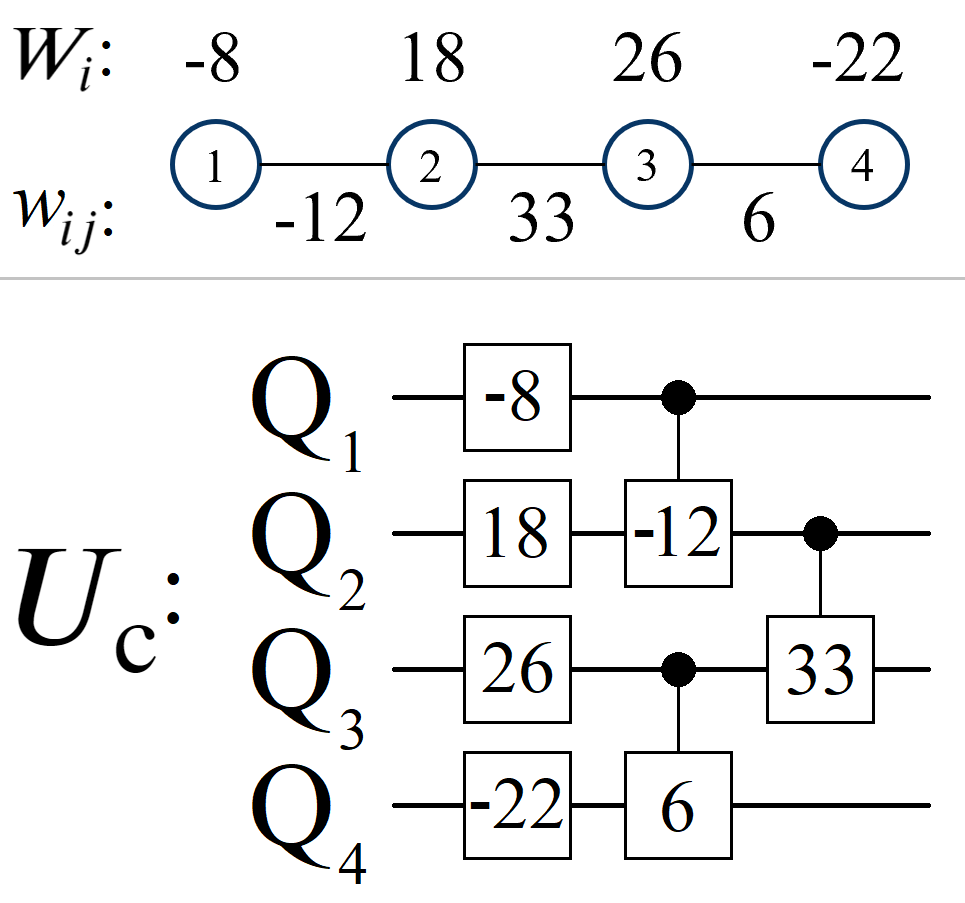}
		\caption{ (top) Example of a $4$-qubit linear QUBO with weighted nodes and edges. (bottom) The same QUBO encoded into a cost oracle $U_{\textrm{c}}$ without scaling.  Each unitary in the circuit is P($\theta$) (single qubit gate) or CP($\theta$) ($2$-qubit gate). }
		\label{Fig.3}
	\end{figure}
	
	Shown above in figure \ref{Fig.3} is an example of a $4$-qubit QUBO cost oracle, where the weighted values $W_i$ and $w_{ij}$ are used as the $\theta$ parameters for the various phase gates.  Although incomplete, we will use this oracle circuit to demonstrate quantum's ability to encode a cost function C(X).  For example, consider the binary solution X$_i = 1101$ and the corresponding quantum basis state $|1101\rangle$.  The classical evaluation of this solution is as follows:
	
	\begin{eqnarray}            
		\textrm{C}(1101) &=&  -8 + 18 - 22 - 12  \nonumber  \\
		&=& -24 \label{Eqn.11}
	\end{eqnarray} 
	
	Now let us compare this to the phase of $|1101\rangle$ after applying $U_{\textrm{c}}$:
	
	\begin{eqnarray}            
		U_{\textrm{c}} |1101 \rangle &=&  \textrm{e}^{i(-8 + 18 - 22 - 12)} |1101 \rangle  \nonumber  \\
		&=& \textrm{e}^{-24 i} |1101 \rangle \label{Eqn.12}
	\end{eqnarray} 
	
	The phase acquired in equation \ref{Eqn.12} is equivalent to the classical evaluation shown in \ref{Eqn.11}, which means that $U_{\textrm{c}}$ is an accurate encoding of C(X).  If we were to now apply $U_{\textrm{c}}$ to the equal superposition state $| \textrm{s} \rangle$ (step 2 in Alg. \ref{Alg.1}), all $2^N$ basis states would receive phases equal to their cost function value.  This is the advantage that quantum has to offer: simultaneously evaluating all possible solutions of a cost function through superposition.  
	
	\subsection{Scaling Parameter $p_{\textrm{s}}$}%
	
	While the cost oracle shown in figure \ref{Fig.3} is capable of reproducing C(X), its use in algorithm \ref{Alg.1} will not yield the optimal solution X$_{\textrm{min}}$ or X$_{\textrm{max}}$.  This is because quantum phases are $2 \pi$ modulo, which is problematic if the numerical scale of C(X) exceeds a range of $2 \pi$.  Consequently if two quantum states receive phases that differ by a multiple of $2 \pi$, then they will both undergo the amplitude amplification process identically.  If this happens unintentionally via $U_{\textrm{c}}$, then our cost oracle cannot be used to minimize or maximize C(X).
	
\begin{figure}[h]            
		\centering
		\includegraphics[scale=.6]{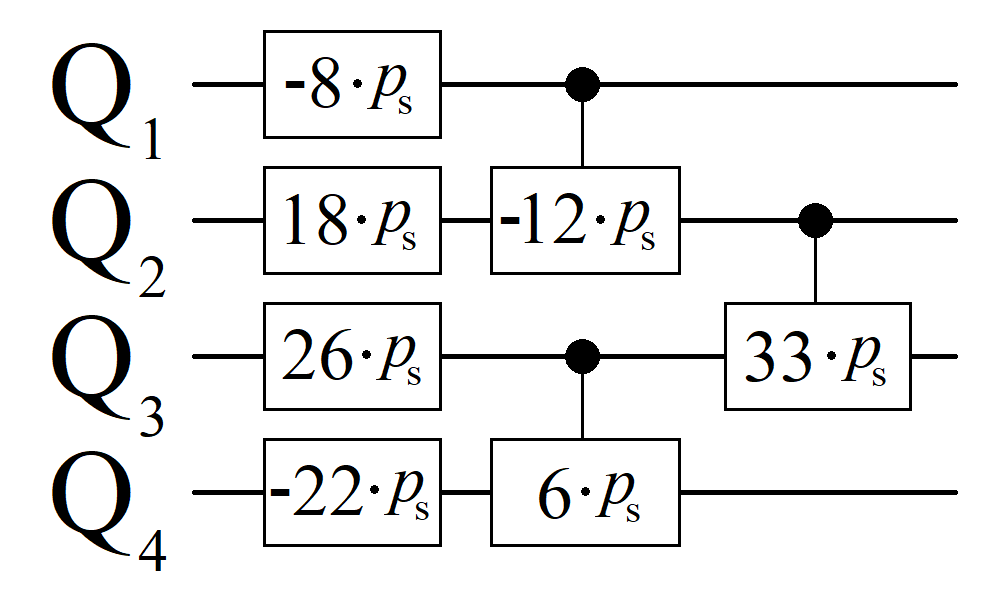}
		\caption{ The $4$-qubit linear QUBO cost oracle $U_{\textrm{c}}$ from figure \ref{Fig.3}, now scaled by $p_{\textrm{s}}$.  }
		\label{Fig.4}
	\end{figure}

	In order to construct $U_{\textrm{c}}$ such that it is usable for amplitude amplification, a scalar parameter $p_{\textrm{s}}$ must be included in all of the phase gates.  The value of $p_{\textrm{s}}$ is problem specific, but its role is always the same: scaling the cumulative phases applied by $U_{\textrm{c}}$ down (or up) to a range where [C(X$_{\textrm{min}}$) , C(X$_{\textrm{max}}$)] is approximately [x , x+$2 \pi$].  This range does not have to be [$0$ , $2 \pi$] exactly, so long as the phases acquired by $| \textrm{X}_{\textrm{min}} \rangle$ and $| \textrm{X}_{\textrm{max}} \rangle$ are roughly $2 \pi$ different.  See figure \ref{Fig.4} below for an example of $p_{\textrm{s}}$ in $U_{\textrm{c}}$'s construction.
	
	Using the scaled oracle shown in figure \ref{Fig.4} above, let us now show how this new $U_{\textrm{c}}$ acts on the basis state $| 1101 \rangle$ from before:
	
	\begin{eqnarray}            
		U_{\textrm{c}} |1101 \rangle &=&  \textrm{e}^{i(-8\cdot p_{\textrm{s}} + 18\cdot p_{\textrm{s}} - 22\cdot p_{\textrm{s}} - 12\cdot p_{\textrm{s}})} |1101 \rangle  \nonumber  \\
		&=&  \textrm{e}^{i(-8 + 18 - 22 - 12) \cdot p_{\textrm{s}}} |1101 \rangle  \nonumber  \\
		&=& \textrm{e}^{-24i \cdot p_{\textrm{s}} } |1101 \rangle \label{Eqn.13}
	\end{eqnarray} 
	
	As shown in equation \ref{Eqn.13} above, multiplying $p_{\textrm{s}}$ into every phase gate has the net effect of scaling the cumulative phase applied by $U_{\textrm{c}}$:  e$^{-24 i} \rightarrow $ e$^{-24i \cdot p_{\textrm{s}}}$.  Note that this is $not$ a global phase, which would have an additive effect on all states rather than a multiplicative one like shown above.
	
	Finding the optimal $p_{\textrm{s}}$ value for boosting X$_{\textrm{min}}$ or X$_{\textrm{max}}$ is non-trivial, and was a major focus of our previous study \cite{koch2}, as well as this one.  In general, the scale of $p_{\textrm{s}}$ needed for finding the optimal solution can be obtained using equation \ref{Eqn.14} below, which scales the numerical range of a problem [C(X$_{\textrm{min}}$) , C(X$_{\textrm{max}}$)] to exactly [x , x+$2 \pi$].  
	
	\begin{eqnarray}            
		p_{\textrm{s}} = \frac{2 \pi}{ \textrm{C}(\textrm{X}_{\textrm{max}}) -  \textrm{C}(\textrm{X}_{\textrm{min}}) }  
		\label{Eqn.14}
	\end{eqnarray} 
	
\begin{figure}[h]            
		\centering
		\includegraphics[scale=.4]{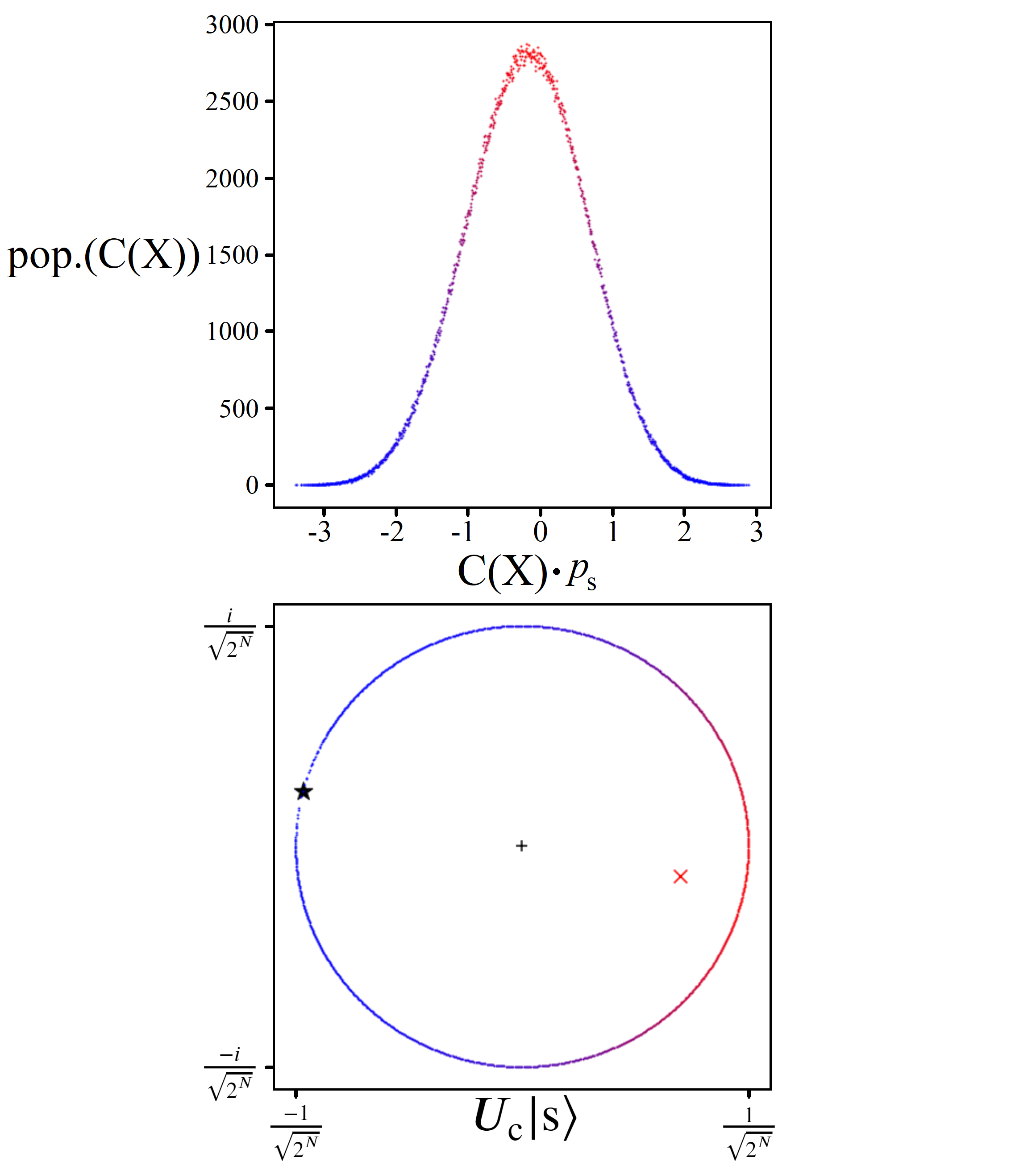}
		\caption{ (top) The $20$-qubit linear QUBO histogram from figure \ref{Fig.2}, scaled by $p_{\textrm{s}}$ according to equation \ref{Eqn.14}.  (bottom)  All $2^{20}$ quantum states after applying $U_{\textrm{c}} | \textrm{s} \rangle$, plotted in amplitude space (the complex plane).  The red-blue color scale shows the density of quantum states in the bottom plot, corresponding to the y-axis of the top histogram.  The states $| \textrm{X}_{\textrm{min}} \rangle$ and $| \textrm{X}_{\textrm{max}} \rangle$ are marked with a black star, the origin a black `+', and average amplitude with a red `x'. }
		\label{Fig.5}
	\end{figure}

	Although equation \ref{Eqn.14} above is guaranteed to solve the $2 \pi$ modulo phase problem mentioned previously, it is almost never the $p_{\textrm{s}}$ value which can be used to find X$_{\textrm{min}}$ or X$_{\textrm{max}}$.  Only in the case of a perfectly symmetric solution space distribution is equation \ref{Eqn.14} the optimal $p_{\textrm{s}}$ value, in which case the states $| \textrm{X}_{\textrm{min}} \rangle$ and $| \textrm{X}_{\textrm{max}} \rangle$ undergo the amplitude amplification process together.  However, realistic optimization problems can be assumed to have a certain degree of randomness or asymmetry to their solution space, producing distributions more akin to figure \ref{Fig.7}.  For this reason, equation \ref{Eqn.14} is better thought of as the starting point for finding the true optimal $p_{\textrm{s}}$, which we discuss later in section IV.B.  For now, equation \ref{Eqn.14} is sufficient for demonstrating $p_{\textrm{s}}$'s role in creating an average amplitude suitable for boosting $| \textrm{X}_{\textrm{min}} \rangle$ or $| \textrm{X}_{\textrm{max}} \rangle$, shown in figure \ref{Fig.5}.
	
	The bottom plot in figure \ref{Fig.5} shows $| \Psi \rangle$ after the first application of $U_{\textrm{c}}$ in algorithm \ref{Alg.1}.  Note the location of the average amplitude (red `x'), which is only made possible by the majority of quantum states which recieve phases near the center of the gaussian in the top plot.  Optimal amplitude amplification occurs when the desired state for boosting is exactly $\pi$ phase different with the mean \cite{boyer,bennett}, which is very close to the situation seen in figure \ref{Fig.5}.  However, since this $U_{\textrm{c}}$ is derived from a QUBO with randomized weights, the $p_{\textrm{s}}$ value provided from equation \ref{Eqn.14} does not exactly produce a $\pi$ phase difference between the optimal states (black star) and the mean amplitude (red `x').  Consequently, the state(s) which does become highly probable from amplitude amplification for this particular $p_{\textrm{s}}$ is not $| \textrm{X}_{\textrm{min}} \rangle$ and $| \textrm{X}_{\textrm{max}} \rangle$, which will be the subject of the coming two sections.

	\section{Gaussian Amplitude Amplification} 
	
	The amplitude space plot depicted at the bottom of figure \ref{Fig.5} is useful for visualizing how a gaussian solution space distribution can be used for boosting, but the full amplitude amplification process is far more complicated.  This is especially true for the QUBOs of this study, which are generated with randomized weights.  Consequently, all of the results which follow throughout the remainder of this study are produced from classical simulations of amplitude amplification using cost oracles derived from linear QUBOs according to equations \ref{Eqn.1} - \ref{Eqn.4}.  For a deeper mathematical insight into these processes, please see \cite{bang,satoh,bench}.
	
	\subsection{Achievable Probabilities}                                                                                                     
	
	Amplitude amplifiation is an appealing quantum algorithm because it solves one of the most fundamental problems of quantum computing: measurement probability.  For example, a single marked state using Grover's oracle with $30$ qubits is capable of achieving a final probability that is only less than $100 \%$ by one billionth of a percent \cite{grover}.  Thus, a natural question to ask when using $U_{\textrm{c}}$ is what kinds of probabilities can it produce for $| \textrm{X}_{\textrm{min}} \rangle$ or $| \textrm{X}_{\textrm{max}} \rangle$?  To answer this we conducted a statistically study of linear QUBOs ranging from length $N = 17$ to $27$.  For each $N$ we generated numerous QUBOs according to equations \ref{Eqn.1} - \ref{Eqn.4}, totals given in appendix $\textbf{A}$.  We then let a classical simulator find the $p_{\textrm{s}}$ value which maximized the probability of measuring $| \textrm{X}_{\textrm{min}} \rangle$ for each QUBO (and for certain cases the optimal $p_{\textrm{s}}$ for $| \textrm{X}_{\textrm{max}} \rangle$ aswell).  Results for each problem size are shown below in figure \ref{Fig.6}.
	
	\begin{figure}[h]            
		\centering
		\includegraphics[scale=.46]{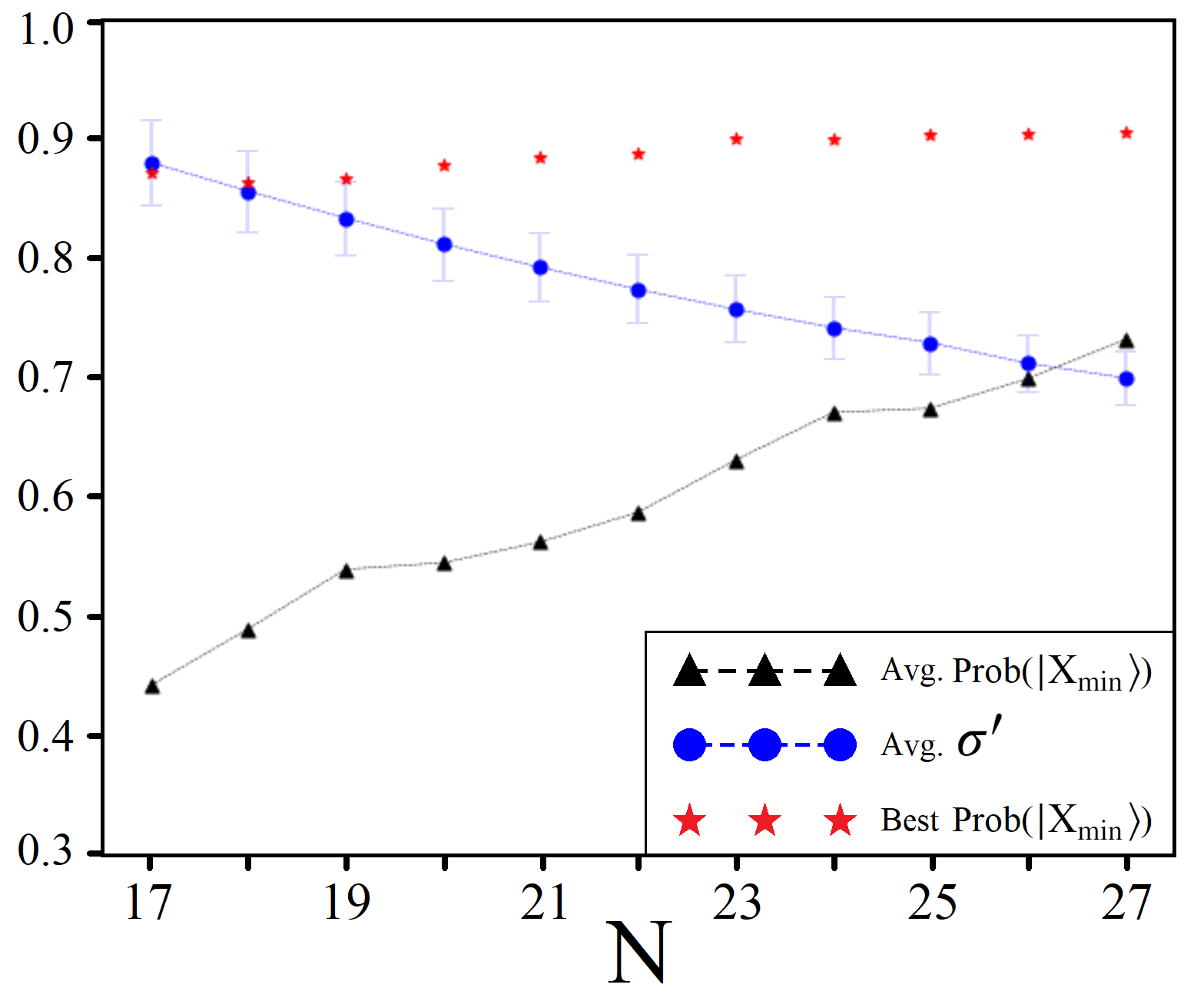}
		\caption{  Results from studying randomly generated linear QUBOs of various sizes $N$.  The number of QUBOs studied per $N$ is provided in appendix $\textbf{A}$.  For each QUBO, the optimal $p_{\textrm{s}}$ value for producing the highest probability of measurement for $| \textrm{X}_{\textrm{min}} \rangle$ was used to record three trends: average probability of $| \textrm{X}_{\textrm{min}} \rangle$ (black triangle), highest recorded probability (red star), and average scaled standard deviation (blue circle).  Error bars showing one standard deviation of each $\sigma$' are provided aswell. }
		\label{Fig.6}
	\end{figure}
	
	\begin{eqnarray}            
		\mu &=& \frac{1}{2^N} \sum_{i} ^{2^N} \textrm{C}(\textrm{X}_i) \label{Eqn.15} \\
		\sigma &=& \sqrt{ \frac{\sum_{i} ^{2^N} ( \textrm{C}(\textrm{X}_i) - \mu )^2}{2^N} } \label{Eqn.16} \\
		\sigma ' &=& \sigma \cdot p_{\textrm{s}} \label{Eqn.17}
	\end{eqnarray} 
	
	Figure \ref{Fig.6} tracks three noteworthy trends found across the various QUBO sizes: the average peak probability achievable for $| \textrm{X}_{\textrm{min}} \rangle$ (black triangle), the highest recorded probability for $| \textrm{X}_{\textrm{min}} \rangle$ (red star), and the average scaled standard deviation $\sigma '$ (blue circle).  For clarity, the derivation of $\sigma '$ is given by equations \ref{Eqn.15} - \ref{Eqn.17}.  This quantity is the standard deviation of a QUBO's solution space distribution after being scaled by $p_{\textrm{s}}$, making it a comparable metric for all QUBO sizes.  In our previous study we demonstrated a result in agreement with figure \ref{Fig.6}, which is the correlation between higher achievable probabilities for $| \textrm{X}_{\textrm{min}} \rangle$ (red star) and smaller scaled standard deviations $\sigma '$ (blue circle) \cite{koch2}.  The latter is what is responsible for increasing the distance between $| \textrm{X}_{\textrm{min}} \rangle$ and the average amplitude like shown in figure \ref{Fig.5}.
	
	\subsection{Solution Space Skewness}                                                                                                     
	
	The relation between $N$, $\sigma '$, and highest prob.($| \textrm{X}_{\textrm{min}} \rangle$) from figure \ref{Fig.6} can be summarized as follows: larger problem sizes tend to produce smaller standard deviations, which in turn lead to better probabilities produced from amplitude amplification.  However, there is a very apparent disconnect between the probabilities capable of each problem size (red stars) versus the average (black triangle).  To explain this, we must first introduce the quantity X$_{\Delta}$ given in equation \ref{Eqn.18} below.
	
	\begin{eqnarray}            
		\textrm{X}_{\Delta}   &=&   2 \mu   - ( \textrm{C}(\textrm{X}_{\textrm{max}})   +  \textrm{C}(\textrm{X}_{\textrm{min}} ) )
		\label{Eqn.18}
	\end{eqnarray} 
	
	\begin{figure}[h]            
		\centering
		\includegraphics[scale=.36]{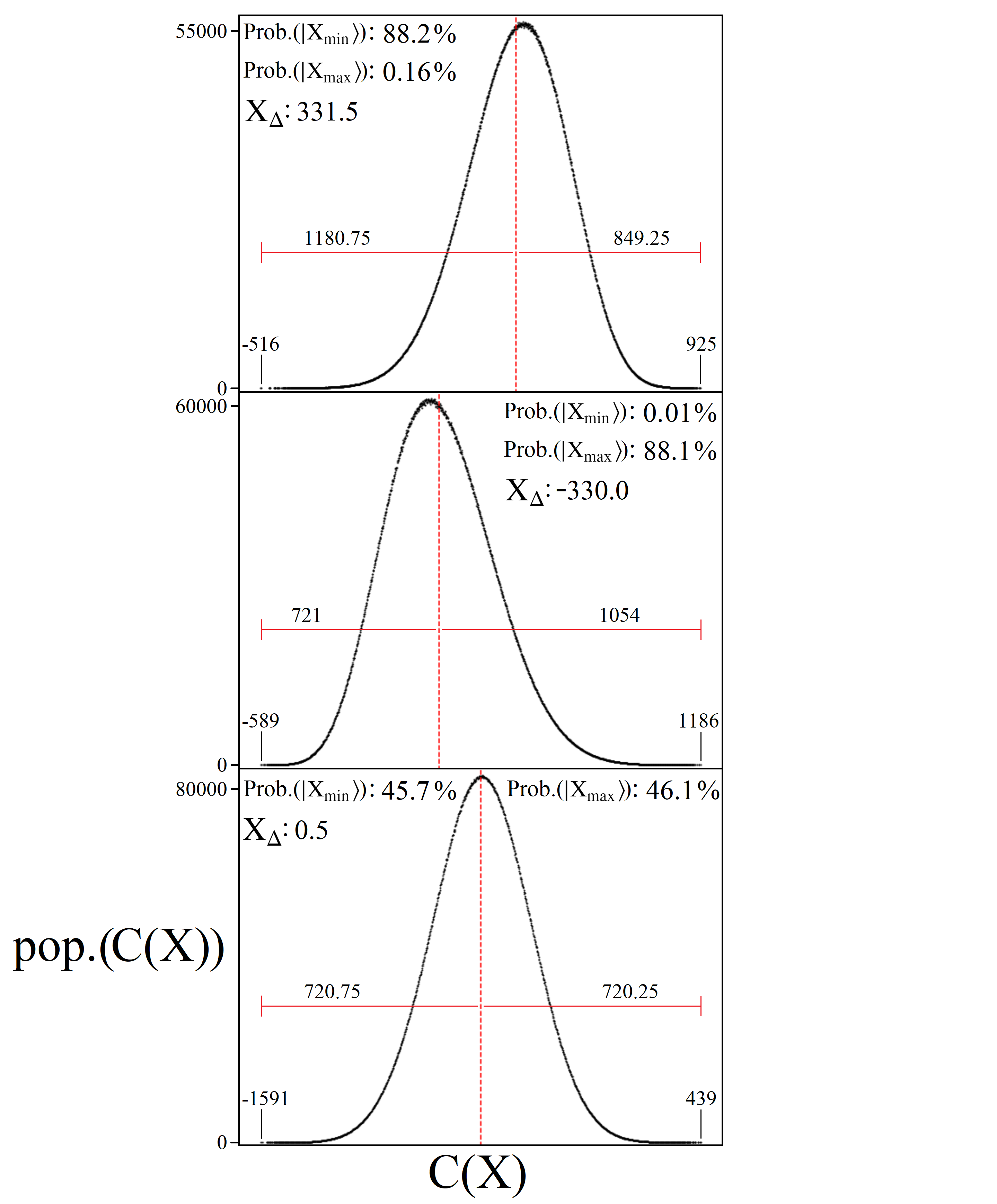}
		\caption{ Three randomly generated QUBO distributions for $N=25$, illustrating X$_{\Delta}$ cases for largely positive (top), largely negative (middle), and near zero (bottom).  In all three plots the exact X$_{\Delta}$ value is reported, as well as the highest achievable probability for $| \textrm{X}_{\textrm{min}} \rangle$ and $| \textrm{X}_{\textrm{max}} \rangle$ (each from a different $p_{\textrm{s}}$ value).  Also shown in each plot are the values for C(X$_{\textrm{min}}$) and C(X$_{\textrm{max}}$), and their numerical distance to the mean $\mu$ (red-dashed line).  }
		\label{Fig.7}
	\end{figure}
	
\begin{figure}[h]            
		\centering
		\includegraphics[scale=.4]{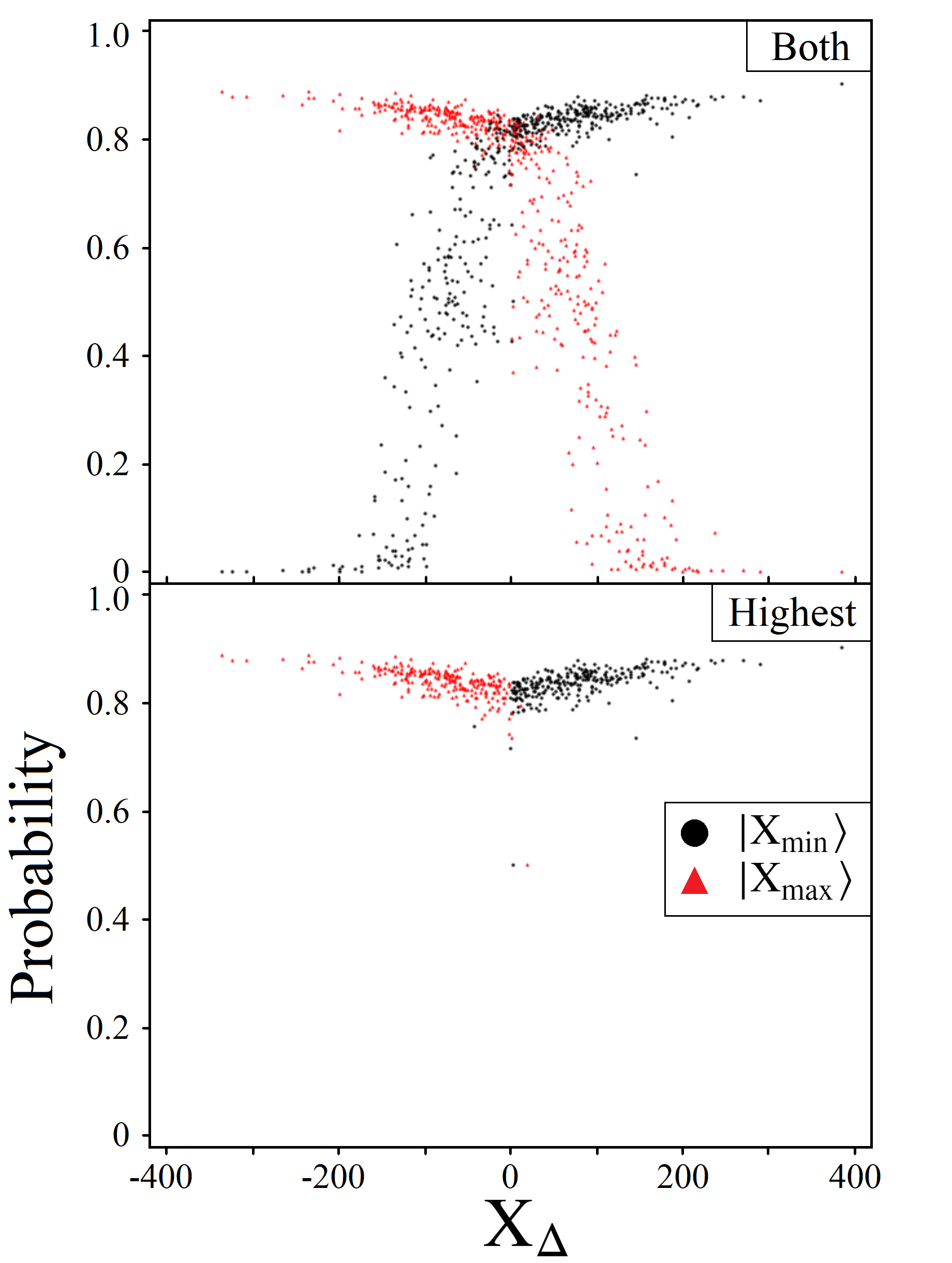}
		\caption{ A total of $1000$ randomly generated linear QUBOs of size $N=23$.  For each QUBO, the highest achievable probability for $| \textrm{X}_{\textrm{min}} \rangle$ (black circle) and $| \textrm{X}_{\textrm{max}} \rangle$ (red triangle) are plotted as a function of X$_{\Delta}$.  The top plot includes both data points per QUBO, while the bottom plot only shows the higher of the two values.}
		\label{Fig.8}
	\end{figure}

	The quantity X$_{\Delta}$ from equation \ref{Eqn.18} is the difference between $\textrm{C}(\textrm{X}_{\textrm{min}})$ and $\mu$ (the mean) minus the difference between $\mu$ and $\textrm{C}(\textrm{X}_{\textrm{max}})$.  A positive value for X$_{\Delta}$ indicates that the mean is closer to $\textrm{C}(\textrm{X}_{\textrm{max}})$, and vice versa for a negative valued X$_{\Delta}$.  In essence, it is a measure of skewness that describes the assymetry of a solution space distribution.  Figure \ref{Fig.7} shows example QUBO distributions for three cases of X$_{\Delta}$, for $N=25$, demonstrating the impact X$_{\Delta}$ has on the ability to boost $| \textrm{X}_{\textrm{min}} \rangle$ versus $| \textrm{X}_{\textrm{max}} \rangle$.  While $\sigma '$ is a strong indicator of a problem's overall aptitude for amplitude amplification, X$_{\Delta}$ determines whether the optimal minimum or maximum solution is boostable, and which is not.  Further evidence of this can be seen in figure \ref{Fig.8}, which shows $1000$ randomly generated linear QUBOs of length $N=23$, and the peak probabilities achievable for $| \textrm{X}_{\textrm{min}} \rangle$ and $| \textrm{X}_{\textrm{max}} \rangle$ as a function of X$_{\Delta}$.

	If we compare the average peak probabilites for $| \textrm{X}_{\textrm{min}} \rangle$ from figure \ref{Fig.6} with the full data of QUBOs shown in figure \ref{Fig.8}, we can see why the average peak probability is significantly lower than the highest recorded.  Across the $1000$ QUBOs studied, it is clear that X$_{\Delta} = 0$ is a dividing point for whether $| \textrm{X}_{\textrm{min}} \rangle$ or $| \textrm{X}_{\textrm{max}} \rangle$ is capable of reaching a significant probability of measurement through amplitude amplification.  For $N=23$, the average prob.($| \textrm{X}_{\textrm{min}} \rangle$) reported in figure \ref{Fig.6} is approximately $64$\%.  However, if instead we only consider QUBOs with X$_{\Delta} > 0$ from figure \ref{Fig.8}, then the average peak probability for $| \textrm{X}_{\textrm{min}} \rangle$ is around $86$\%, and likewise for $| \textrm{X}_{\textrm{max}} \rangle$ when X$_{\Delta} < 0$.
	
	Together, figures \ref{Fig.7} and \ref{Fig.8} demonstrate the significance of knowing X$_{\Delta}$ from an experimenter's perspective.  Depending on the optimization problem of interest, it is reasonable to assume that an experimenter may be interested in finding only X$_{\textrm{min}}$ or X$_{\textrm{max}}$.  But without any $\textit{a}$ $\textit{priori}$ knowledge of a problem's underlying solution space, specifically X$_{\Delta}$, the experimenter may unknowingly be searching for a solution which is probabilistically near impossible to find through amplitude amplification.  For example, consider the QUBO distribution illustrated in the top plot of figure \ref{Fig.7}, and the peak probability for boosting $| \textrm{X}_{\textrm{max}} \rangle$: $0.16$\%.  Although it is ideal to have insight into a particular problem's X$_{\Delta}$ before using amplitude amplification, as we demonstrate in section V., information about X$_{\Delta}$ can be inferred through measurement results.

	\subsection{Sampling for $p_{\textrm{s}}$ }                                                                                                     

	If a particular optimization problem is suitable for amplitude amplification, then the speed of the quantum algorithm outlined in this study is determined by how quickly the optimal $p_{\textrm{s}}$ value can be found.  Here we shall show that sampling a cost function C(X) can provide reliable information for approximating $p_{\textrm{s}}$ from equation \ref{Eqn.14}, which can then be used to begin the variational approach outlined in sections V. and VI.  Importantly, the number of cost function evaluations needed is significantly less than either a classical or quantum solving speed.  The strategy outlined in equations \ref{Eqn.19} - \ref{Eqn.29} below can be used for approximating $p_{\textrm{s}}$ when the experimenter is expecting an underlying solution space describable by a gaussian function (equation \ref{Eqn.6}).  If another type of distribution is expected, then the function used in equation \ref{Eqn.22} could in principle be modified accordingly (for example, sinusoidal, polynomial, exponential \cite{bench}).  
	
	Suppose we sample a particular cost function C(X) $M$ times, where $M << 2^N$.  We will define the set $\mathbb{M}$ as the collection of values C(X$_i$) obtained from these samples.
	
	\begin{eqnarray}            
		\mathbb{M} = \{ \textrm{C}(\textrm{X}_1) , \textrm{C}(\textrm{X}_2) , ... , \textrm{C}(\textrm{X}_M) \} \label{Eqn.19} 
	\end{eqnarray} 
	
	Using these $M$ values, we can compute an approximate mean and standard deviation.
	
	\begin{align}
		\tilde{\mu} &= \frac{1}{M} \sum_{c\in \mathbb{M}} c \\
		\tilde{\sigma} &= \sqrt{\frac{ \sum_{c\in \mathbb{M}} {\left(c - \tilde{\mu}\right)}^{2}}{M}}
	\end{align}
	
	In order to use equation \ref{Eqn.14} for obtaining $p_{\textrm{s}}$, we need approximations for C(X$_{\textrm{min}}$) and C(X$_{\textrm{max}}$).  If we assume an underlying gaussian structure to the problem's solution space, then we can write down the following equation to describe it:
	
	\begin{eqnarray}            
		2^N &=& \int_{-\infty}^{\infty} \tilde{\alpha} \hspace{0.02cm} \textrm{e}^{\frac{(x - \tilde{\mu})^2}{2 \tilde{\sigma}^2}} dx \label{Eqn.22} \\ 
		&=& - \tilde{\alpha} \hspace{0.04cm} \tilde{\sigma} \sqrt{\frac{\pi}{2}}  \textrm{erf} \left(  \frac{\tilde{\mu} - x}{\sqrt{2} \tilde{\sigma}} \right)_{-\infty}^{\infty}  \label{Eqn.23} \\ 
		&=& - \tilde{\alpha} \hspace{0.04cm} \tilde{\sigma} \sqrt{\frac{\pi}{2}} \cdot [ -1 - 1 ] \label{Eqn.24}
	\end{eqnarray}
	
	where erf() is the gaussian error function.  Using equation \ref{Eqn.24}, we can rearrange terms and solve for an approximation to the height of the gaussian.
	
	\begin{eqnarray}            
		\tilde{\alpha}	= \frac{2^{N-1}}{\tilde{\sigma} \sqrt{\frac{\pi}{2}}} \label{Eqn.25}
	\end{eqnarray}
	
	With the values $\tilde{\mu}$, $\tilde{\sigma}$, and $\tilde{\alpha}$ obtained from sampling, we can now approximate C(X$_{\textrm{min}}$) and C(X$_{\textrm{max}}$) using equation \ref{Eqn.26} below.
	
	\begin{eqnarray}            
		\tilde{\textrm{G}}(x)	=  \tilde{\alpha} \textrm{e}^{\frac{(x - \tilde{\mu})^2}{2 \tilde{\sigma}^2}} = 1 \label{Eqn.26}
	\end{eqnarray}
	
	Solving for $x$ yields the following two values:
	
	\begin{eqnarray}            
		x_{\pm} = \tilde{\mu} \pm \tilde{\sigma} \sqrt{ -2 \textrm{ln} \left(   \frac{ 1 }{\tilde{\alpha}}   \right) } \label{Eqn.27}
	\end{eqnarray}
	
	which can be expressed in terms of the two quantities originally derived from sampling:
	
	\begin{eqnarray}            
		x_{\pm} = \tilde{\mu} \pm \tilde{\sigma} \sqrt{ -2 \textrm{ln} \left(   \frac{ \tilde{\sigma} \sqrt{ \pi / 2 } }{2^{N-1}}   \right) } \label{Eqn.28}
	\end{eqnarray}
	
	And finally, the solutions $x_{\pm}$ can be used to obtain $p_{\textrm{s}}$.
	
	\begin{eqnarray}            
		\tilde{p}_{\textrm{s}} = \frac{2 \pi}{ x_{+} - x_{-} }\label{Eqn.29}
	\end{eqnarray}
	
	The reason we set equation \ref{Eqn.26} equal to $1$, and the integral in equation \ref{Eqn.22} equal to $2^N$, is because $\tilde{\textrm{G}}$($x$) is modeling the histogram of a QUBO's solution space, like shown in figure \ref{Fig.2}.  This means that the total number of solutions to C(X) is $2^N$, and similarly the minimum number of distinct C(X$_i$) solutions for a given cost function is $1$.  Therefore, after setting the integral in equation \ref{Eqn.22} equal to $2^N$, solving $\tilde{\textrm{G}}$($x$)$=1$ yields approximations for C(X$_{\textrm{min}}$) and C(X$_{\textrm{max}}$) on the tails of the gaussian.  
	
	To demonstrate how well sampling is able to approximate equation \ref{Eqn.14}, we tested the strategy outlined above against the $1000$ QUBOs from figure \ref{Fig.8} ($N=23$).  For four values of $M$: $100$, $500$, $1000$, and $2000$, each QUBO was used for $50$ trials of random sampling to produce approximate $\tilde{p}_{\textrm{s}}$ values.  These values were then compared to the true value of $p_{\textrm{s}}$ from equation \ref{Eqn.14}, as given by equation \ref{Eqn.30} below, and finally averaged together to produce table \ref{Tab.1}.
	
	\begin{eqnarray}            
		\tilde{p}_{\textrm{s}} \hspace{0.1cm} \textrm{Error}  = \frac{| \tilde{p}_{\textrm{s}} - p_{\textrm{s}}| }{ p_{\textrm{s}}}     \label{Eqn.30} 
	\end{eqnarray}

	\begin{table}[h]
		\begin{tabular}{|c|c|c|c|c|}
			\hline
			$M$   & 100 & 500 & 1000 &  2000   \\ \hline
			$\hspace{0.05cm}$ Average $\tilde{p}_{\textrm{s}}$ Error $\hspace{0.05cm}$   &   $\hspace{0.1cm}$ 7.28\% $\hspace{0.1cm}$& $\hspace{0.1cm}$ 6.37\% $\hspace{0.1cm}$ & $\hspace{0.1cm}$ 6.31\% & $\hspace{0.1cm}$ 6.29\%   $\hspace{0.1cm}$  \\ \hline
		\end{tabular}
		\caption{ Average error in approximating $p_{\textrm{s}}$ using equations \ref{Eqn.19} - \ref{Eqn.29}. Each value comes from $50$,$000$ independent sampling trials on linear QUBOs of size $N=23$.}
		\label{Tab.1}
	\end{table}
	
	The significant result from table \ref{Tab.1} is that sampling $100$ - $500$ times, on a cost function of $2^{23}$ solutions, is accurate enough to produce an approximate $\tilde{p}_{\textrm{s}}$ value with an expected error of only $7$\%.  And as we show in the next section, this is enough accuracy to use for either a heuristic or variational approach for finding optimal solutions.
	
	\section{Variational Amplitude Amplification} 
	
	The results of sections II - IV. demonstrate quantum's aptitude for encoding and solving a QUBO problem using amplitude amplification.  In this section we discuss how this potential can be realized from an experimental perspective.  In particular, we focus on amplitude amplification's ability to find optimal solutions under realistic circumstances with limited information.  The results of this section are then used to motivate section VI., in which we discuss how amplitude amplification can be used in a hybrid classical-quantum model of computing, similar to other successful variational approaches \cite{qaoa,qaoa2,vqe}.

	\subsection{Boosting Near-Optimal Solutions}%
	
	The results shown in figures \ref{Fig.6} - \ref{Fig.8} focus on quantum's potential for finding $| \textrm{X}_{\textrm{min}} \rangle$ and $| \textrm{X}_{\textrm{max}} \rangle$, the optimal solutions which minimize/maximize a given cost function C(X).  However, in order to understand how amplitude amplification can be used in a variational model, it is equally as important that non-optimal $| \textrm{X}_{i} \rangle$ states are also capable of boosting.
	
\begin{figure}[h]            
		\centering
		\includegraphics[scale=.4]{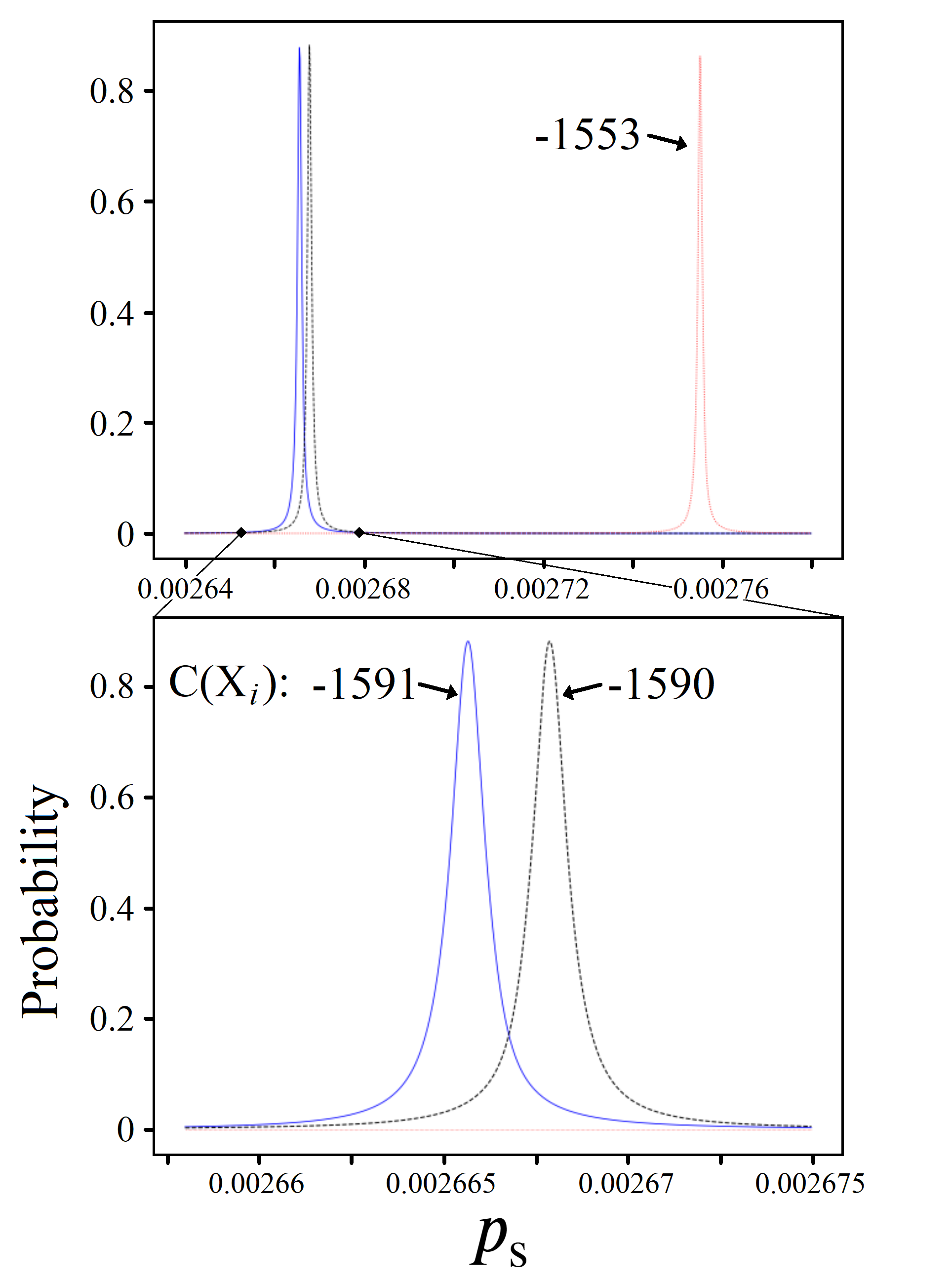}
		\caption{ Plots of $| \textrm{X}_i \rangle$ state probability from amplitude amplification as a function of $p_{\textrm{s}}$, for $| \textrm{X}_{\textrm{min}} \rangle$ (blue-solid) and the next two minimal solutions (black and red-dashed).  Cost function values C(X$_{i}$) are reported next to each state's plot, corresponding to the QUBO from the top plot in figure \ref{Fig.7}.  The bottom plot is a zoomed in scale of the top plot, depicting the same data points.}
		\label{Fig.9}
	\end{figure}

	As discussed in the conclusion of our previous study \cite{koch2}, as well as sections III.C and IV.C, the most difficult aspect of using algorithm Alg.\ref{Alg.1} from an experimental standpoint is finding $p_{\textrm{s}}$.  More specifically, finding an optimal $p_{\textrm{s}}$ for boosting $| \textrm{X}_{\textrm{min}} \rangle$ or $| \textrm{X}_{\textrm{max}} \rangle$ is a challenge due to the limited amount of information that one can learn through measurements alone.  An example of this can be seen in figure \ref{Fig.9}, which shows the peak achievable probabilities of the three lowest $| \textrm{X}_i \rangle$ states as a function of $p_{\textrm{s}}$ ($| \textrm{X}_{\textrm{min}} \rangle$ and the next two minimum solutions), for the QUBO corresponding to X$_{\Delta} = 331.5$ from figure \ref{Fig.7}.

	The challenge presented in figure \ref{Fig.9} is the narrow range of $p_{\textrm{s}}$ values for which each $| \textrm{X}_{i} \rangle$ state is able to achieve meaningful probabilities of measurement.  From an experimental perspective, the $p_{\textrm{s}}$ regions outside these bands are only capable of producing quantum superposition states which are slightly better than $| \textrm{s} \rangle$, the equal superposition starting state. Thus, an experimenter could use a $p_{\textrm{s}}$ value that is incredibly close to optimal, but only see seemingly random measurement results through repeat implementations of Alg.\ref{Alg.1}.
	
	Our proposed solution to the $p_{\textrm{s}}$ problem as described above is twofold: 1) We must widen our view of useful $p_{\textrm{s}}$ values and see where other $| \textrm{X}_{i} \rangle$ states become highly probable, and 2) put less burden on quantum to find optimal solutions alone when an assisting classical approach may be more suitable.  In this section we focus on addressing (1), which will then motive (2) in section VI.
	
	Suppose we aren't solely interested in using quantum to find the exact optimal solution C(X$_{\textrm{min}}$), but instead are content with any X$_i$ within the best $50$ answers ($50$ lowest C(X) values).  In order for amplitude amplification to be viable in this heuristic context, it requires significant probabilities of measurement for these non-optimal solution states, similar to figure \ref{Fig.9}.  Additionally, an experimenter must be able to quickly and reliably find the $p_{\textrm{s}}$ values which produce them.  Shown below in figure \ref{Fig.10} is a plot which provides insight into the feasibility of both of these questions, for the QUBO corresponding to figure \ref{Fig.9}.
	
	\begin{figure}[h]            
		\centering
		\includegraphics[scale=.4]{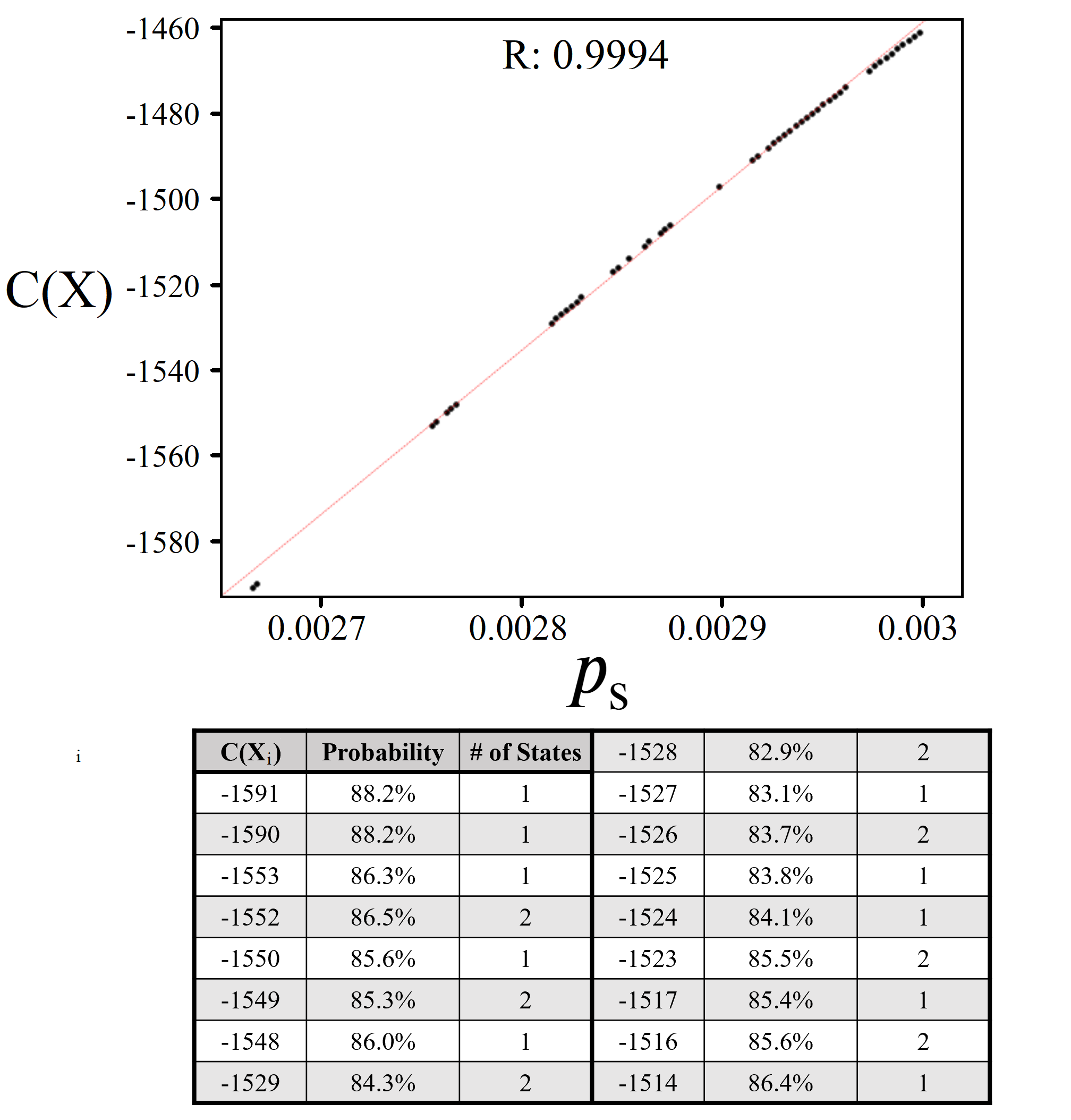}
		\caption{ (top) A plot of the $50$ lowest C(X$_{i}$) values as a function of $p_{\textrm{s}}$, for the X$_{\Delta} = 331.5$ QUBO from figure \ref{Fig.7}.  Each data point represents the $p_{\textrm{s}}$ value where the $| \textrm{X}_i \rangle$ state(s) is most probable.  A linear regression best-fit is shown by the red-dotted line, with its R correlation value reported at the top (equation \ref{Eqn.B5} from appendix $\textbf{B}$). (bottom) A table of values for the $20$ best solutions.  Each entry reports: the cost function value C(X$_{i}$), the peak probability for the $| \textrm{X}_i \rangle$ state(s), and the number of unique X$_i$ solutions that result in the same C(X$_{i}$) value.}
		\label{Fig.10}
	\end{figure}
	
	Figure \ref{Fig.10} shows the full $p_{\textrm{s}}$ range for which an experimenter could find the $50$ best solutions for minimizing C(X) via quantum measurements.  The black circles indicate on the x-axis the $p_{\textrm{s}}$ values where each $| \textrm{X}_{i} \rangle$ state (or multiple states) is maximally probable, aligning with its corresponding C(X$_{i}$) value along the y-axis.  Numeric values for peak probabilities of the best $20$ solutions are provided in the table below the plot, as well as a linear regression best fit (red-dotted line) for the overall $50$ data points.  The reported R correlation value is given by equation \ref{Eqn.B5} in appendix $\textbf{B}$.
	
	There are several significant results displayed in figure \ref{Fig.10}, the first of which requires returning to equation \ref{Eqn.2}.  By limiting the allowed weighted values for $W_i$ and $w_{ij}$ to integers, all solutions to C(X) are consequently integers as well.  This means that the linear correlation shown in the figure can in principle be used to predict $p_{\textrm{s}}$ values where integer C(X$_i$) solutions must exist.  If $W_i$ and $w_{ij}$ are instead allowed to take on float values, which is more general of realistic optimization problems, the linearity of solutions like shown still persists but cannot be used for predictions of allowed C(X) values as reliably.
	
	The linear best fit shown in figure \ref{Fig.10} is accurate for the top $50$ solutions, but extending the $p_{\textrm{s}}$ scale further reveals that it is only an approximation applicable to a small percentage of states.  This is shown in figure \ref{Fig.11} below, which once again is a $p_{\textrm{s}}$ vs. C(X) plot for the same QUBO, but now for the best $400$ minimizing solutions.  It is clear from the data in this figure that the top $400$ solutions are in no way linearly aligned, which is a more expected result given the complicated nature of these imperfect gaussian distributions undergoing amplitude amplification.  However, although the data is not linear, there is very clearly a curved structure that could be utilized to predict $p_{\textrm{s}}$ values in the same manner described above.

	\begin{figure}[h]            
		\centering
		\includegraphics[scale=.4]{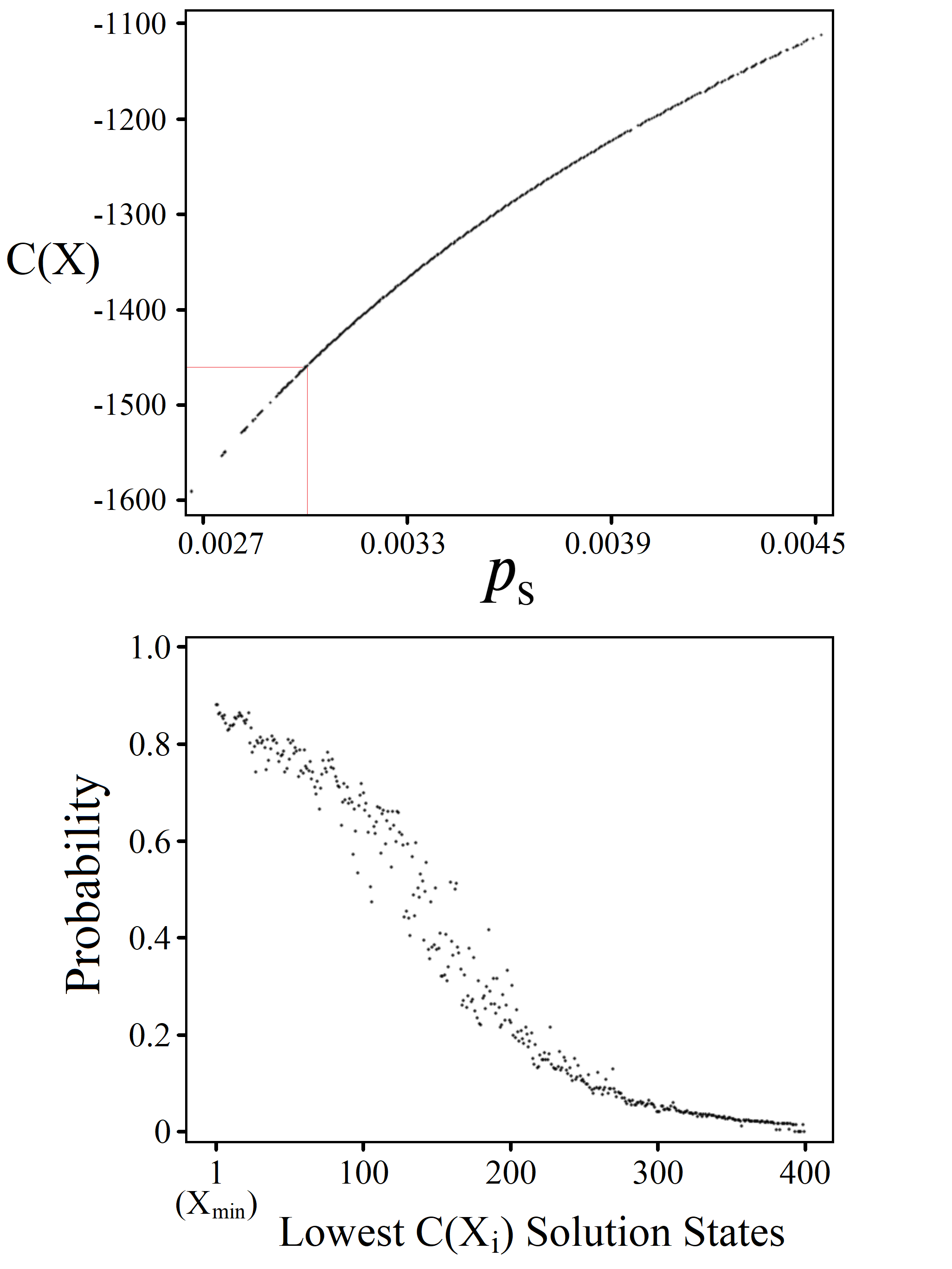}
		\caption{ (top) A plot of the $400$ lowest C(X$_{i}$) values as a function of $p_{\textrm{s}}$, for the X$_{\Delta} = 331.5$ QUBO from figure \ref{Fig.7}.  Each data point represents the $p_{\textrm{s}}$ value where the $| \textrm{X}_i \rangle$ state(s) is most probable.  The red box in the lower left corner represents the $p_{\textrm{s}}$ region depicted in figure \ref{Fig.10}. (bottom)  The probabilities achieved for these $400$ lowest $| \textrm{X}_i \rangle$ states using the $p_{\textrm{s}}$ values shown in the top plot.  Each state is plotted in order of it's rank from $1$ (X$_{\textrm{min}}$) to $400$ ($400^{\textrm{th}}$ lowest C(X$_{i}$) solution).  }
		\label{Fig.11}
	\end{figure}

	\begin{figure*}[t]               
		\centering
		\includegraphics[scale=.34]{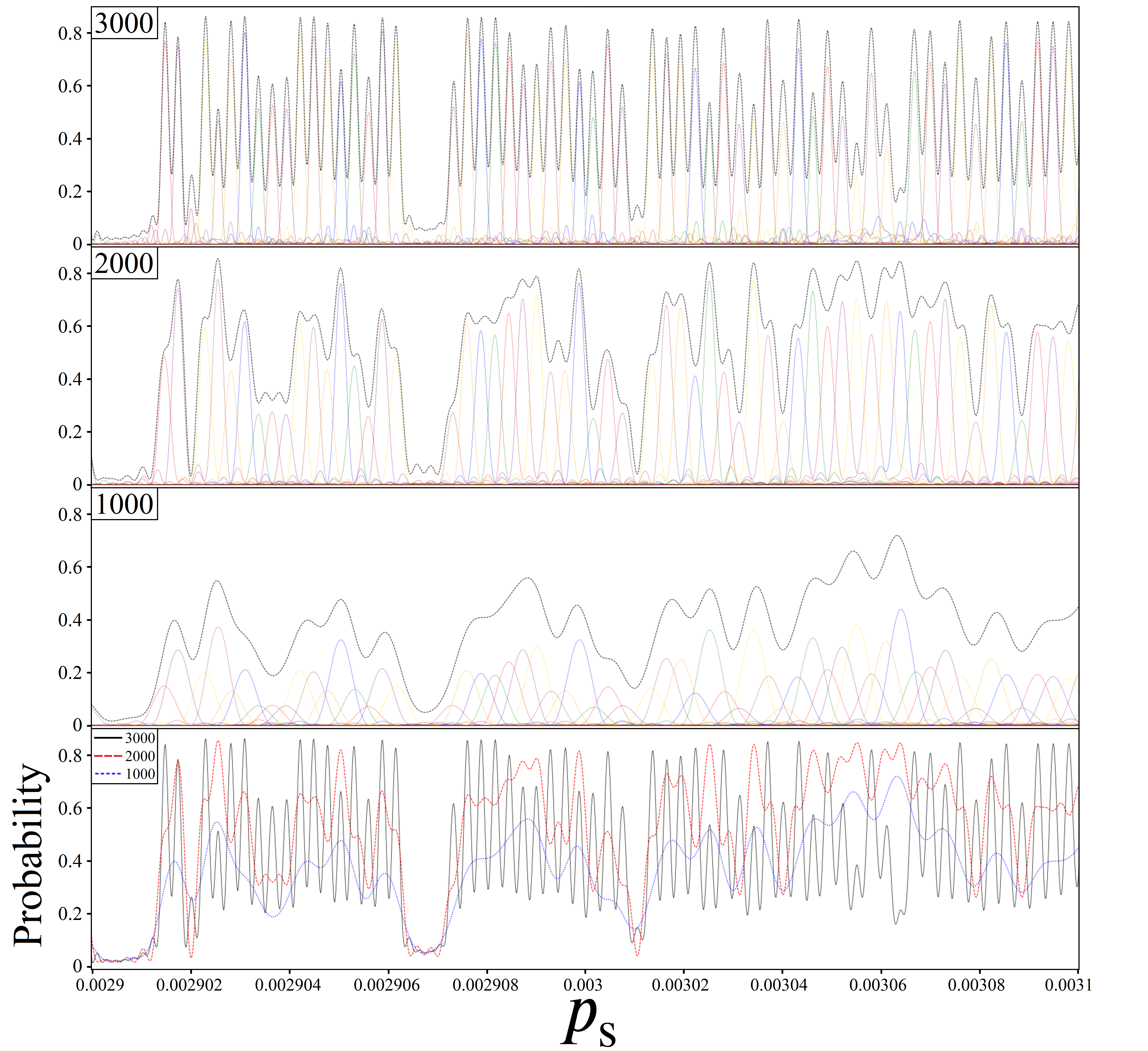}
		\caption{ Plots of $| \textrm{X}_{i} \rangle$ state probabilities as a function of $p_{\textrm{s}}$, for the $N=25$ QUBO shown in figures \ref{Fig.10} and \ref{Fig.11}.  The top three panels show individual state probabilities as solid-colored lines, for three different constant $k$ iterations ($1000$, $2000$, and $3000$) across the $p_{\textrm{s}}$ region depicted on the x-axis.  An additional black-dashed line is also shown, which records the cumulative probability of the five most probable solutions $| \textrm{X}_{i} \rangle$ at any given $p_{\textrm{s}}$ value.  These cumulative probabilities are also replotted in the bottom most panel for comparison. }
		\label{Fig.12}
	\end{figure*}

	It is important to note that in both figures \ref{Fig.10} and \ref{Fig.11}, the manner in which the solution states $|\textrm{X}_i\rangle$ are found to be most probable is sequential.  This means that if a particular state $|\textrm{X}_i\rangle$ is most probable at a certain value $p_{\textrm{s}} = x$, all solutions C(X$_j$) $<$ C(X$_i$) will have peak probabilities at values $p_{\textrm{s}} < x$.  However, the bottom plot in figure \ref{Fig.11} shows that the further a solution state is from $| \textrm{X}_{\textrm{min}}\rangle$ the lower its achievable peak probability.  This means that there is a limit to how many solutions are viable for amplitude amplification to find.  As we discuss in the coming subsections, these are the key underlying features that we must consider when constructing a variational amplitude amplification algorithm.

	\subsection{Constant Iterations}%
	
	In order to construct an algorithm which capitalizes on the structure and probabilities shown in figure \ref{Fig.11}, we must consider an additional piece of information not illustrated in the figure: step $3$ of Alg. \ref{Alg.1}, iterations $k$.  The data points in the figure are indeed the $p_{\textrm{s}}$ values which produce the highest probabilities of measurements, but unfortunately they are achieved using different iteration counts.  In principle this means that an experimenter must decide both $p_{\textrm{s}}$ and $k$ before each amplitude amplification attempt, further complicating the information learned from measurement results.
	
	Unlike $p_{\textrm{s}}$, which is difficult to learn because it depends on the collective $2^N$ solutions to C(X), approximating a good iteration count $k$ is easier.  It turns out that the standard number of Grover iterations $ k_{\textrm{G}} = \frac{\pi}{4}\sqrt{N/M}$, where $N$ is the total number of quantum states and $M$ is the number of marked states, is equally applicable when using $U_{\textrm{c}}$ as well.  If an experimentor can use $k \approx k_{\textrm{G}}$ iterations for a cost oracle $U_{\textrm{c}}$ and find significant probabilities of measurment, then a variational amplitude amplification strategy can be reduced to a single parameter problem: $p_{\textrm{s}}$.  Figure \ref{Fig.12} demonstrates that this is indeed viable, showcasing $| \textrm{X}_{i} \rangle$ solution state probabilities as a function of $p_{\textrm{s}}$ for three different choices of $k$.
	
	The QUBO corresponding to figure \ref{Fig.12} is the same $N=25$ example for figures \ref{Fig.10} and \ref{Fig.11}.  For instances where multiple states correspond to the same numerical solution (C(X$_i$) $=$ C(X$_j$)), the solid-color line shown represents their joint probability: Prob.( $| \textrm{X}_i \rangle $ ) + Prob.( $| \textrm{X}_j \rangle$ ) (note that these individual probabilities are always equal).  Examples of this can also be seen in the table included in figure \ref{Fig.10}. Additionally, a black-dashed line is shown in the top three plots, tracking the collective probability of the five most probable solutions at any given $p_{\textrm{s}}$.  These three lines are then replotted in the bottom panel for comparison.
	
	The $p_{\textrm{s}}$ region shown in figure \ref{Fig.12} was chosen to illustrate a scenario where variational amplitude amplification is most viable.  For $p_{\textrm{s}} > 0.00291$, nearly every possible integer solution C(X$_{i}$) $\geq -1497$ exists via some binary combination for this particular QUBO problem.  The exceptions where certain integer solutions do not exist can be seen clearly in the $p_{\textrm{s}}$ regions with very low probability, for example $ 0.0029065 \leq p_{\textrm{s}}  \leq 0.002907$.  Contrast to the region shown in this figure, once $p_{\textrm{s}}$ becomes closer to where $| \textrm{X}_{\textrm{min}} \rangle$ is maximally probable, then measurment probabilities become more akin to figure \ref{Fig.9}.  Thus, it is more strategic for a hybrid algorithm to start in a $p_{\textrm{s}}$ region like figure \ref{Fig.12}, where measurement results can consistently yield useful information.
	
	\subsection{Information Through Measurements}%
	
	From an experimental perspective, a significant result from figure \ref{Fig.12} are the black-dashed lines shown in the top three plots.  At $k = 3000$ ($k_{\textrm{G}} \approx 4500$ for 25 qubits, $M=1$), the black-dashed line is almost entirely composed of the single most probable solution state(s).  With probabilities around $70-80$\% for many of the states shown, it is realistic that the same $| \textrm{X}_i \rangle$ state could be measured twice in only $2-4$ amplitude amplification attempts.  Two measurements yielding the same C(X$_i$) value (possibly from different X$_i$) is a strong experimental indicator that the $p_{\textrm{s}}$ value used is very close to optimal for that solution, corresponding to the data points from figures \ref{Fig.10} and \ref{Fig.11}.  Confirming $3-4$ different data points in this manner can then be used to approximate the underlying curved structure of these figures, which in turn could be used to predict $p_{\textrm{s}}$ values where $| \textrm{X}_{\textrm{min}} \rangle$ may exist.
	
	While using $k$ closer to $k_{\textrm{G}}$ is good for getting the maximal probability out of solution states, the $k=1000$ and $2000$ plots in figure \ref{Fig.12} support a different strategy for quantum.  At $k=2000$, the black-dashed line is still primarily composed of the single most probable $| \textrm{X}_i \rangle $ state(s), but critically it does not have the same dips in probability between neighboring solutions.  Instead, the cumulative probability stays just as high for these in-between $p_{\textrm{s}}$ regions, sometimes even higher!  If we now look at the $k=1000$ plot, this trend becomes even more prevelant, whereby the cumulative probability plot is on average $20-30$\% higher than any individual $| \textrm{X}_i \rangle$ state.  Interestingly, the bottom panel of figure \ref{Fig.12} shows that cumulative probability plot for $k=1000$ is higher than the $k=3000$ line in many regions.  Thus, if the role of quantum is to simply provide a heuristic answer \cite{durand}, not necessarily $| \textrm{X}_{\textrm{min}} \rangle$, then using lower $k$ values is favorable for a few reasons.  Firstly, we can anticipate solutions in a $p_{\textrm{s}}$ region where multiple states share the same cost function value, so one can expect $M > 1$ more frequently when using $ k_{\textrm{G}} = \frac{\pi}{4} \sqrt{N/M}$.  Secondly, the amplitude amplification process itself is faster due to smaller $k$, which makes it more achievable on noisy qubits due to shallower circuit depths.
	
\begin{figure}[h]            
		\centering
		\includegraphics[scale=.44]{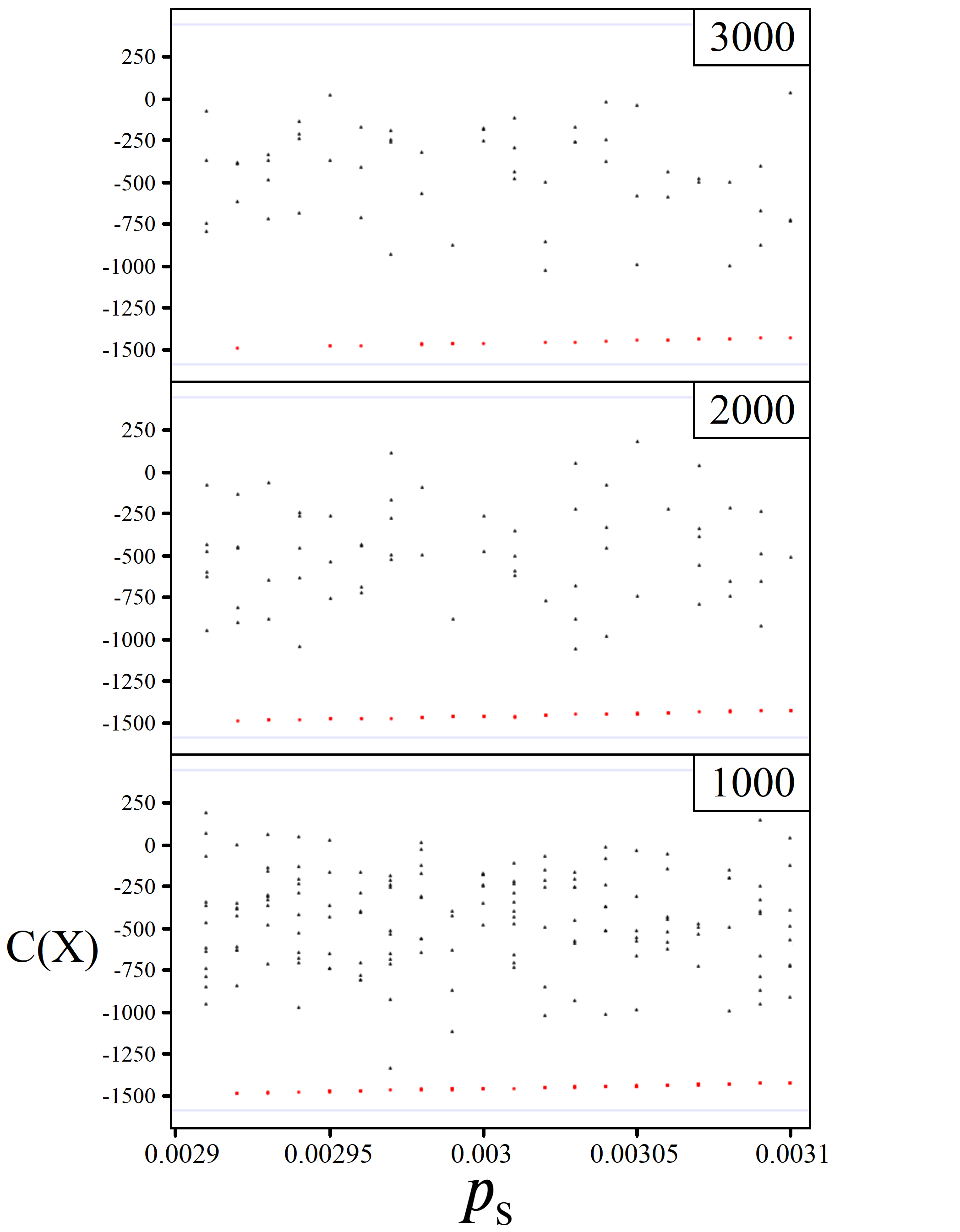}
		\caption{ Simulated measurement results corresponding to the probabilities shown in figure \ref{Fig.12}, produced by amplitude amplification for various values of $p_{\textrm{s}}$ (x-axis) and $k$ ($1000$, $2000$, and $3000$).  At each of the $p_{\textrm{s}}$ values simulated, the number of measurements per experiment $t$ was chosen based on $k$ as follows ($t$,$k$): ($4$,$3000$) , ($6$,$2000$) , ($12$,$1000$), such that $t \cdot k = 12000$.  Measurement results which yielded C(X$_i$)$< -1350$ are plotted as red circles, otherwise as black triangles.  Blue lines for C(X$_{\textrm{min}}$) and C(X$_{\textrm{max}}$) are plotted as well.    }
		\label{Fig.13}
	\end{figure}

	The optimal use of $k$ is a non-trivial challenge to an experimentor.  However, as illustrated in figure \ref{Fig.12}, amplitude amplification can still be effective with a wide range of different $k$ values.  To further demonstrate this, figure \ref{Fig.13} shows three plots of simulated measurements over the $p_{\textrm{s}}$ range depicted in figure \ref{Fig.12}.  Using the $k$ values $1000$, $2000$, and $3000$, each plot shows data points representing probabilistic measurements at regular intervals of $p_{\textrm{s}}$.  In order to compare the $k$ value's effectiveness more equally, the number of measurements taken per $p_{\textrm{s}}$ value, $t$, was chosen such that $t \cdot k = 12000$ is consistent across all three experiments.  Thus, each of the three plots in figure \ref{Fig.13} represents the same total number of amplitude amplification iterations divided among $t$ experimental runs. 
	
	The data points shown in figure \ref{Fig.13} are separated into two categories, which are easily recognizable from an experimental perspective.  Measurements which yielded  C(X$_i$) $< -1350$ are plotted as red circles, while all other measurements are plotted as black triangles.  As illustrated for all three values of $k$, the red data points can be seen as producing near linear slopes, all of which would signal to the experimenter that these measurement results are leading to X$_{\textrm{min}}$.  The motivation for figure \ref{Fig.13} is to demonstrate that the same underlying information can be experimentally realized using different $k$ values.  Thus, when to use $k=3000$ versus $k=1000$ is a matter of optimization, which we discuss in section VI. as the role of a classical optimizer for a hybrid model.
	
	\subsection{Quantum Verification}%
	
	The results of the previous subsections demonstrate the capacity for amplitude amplification as a means for finding a range of optimal X$_i$ solutions.  However, regardless of whether these solutions are found via quantum or classical, a separate problem lies in verifying if a given solution is truly the global minimum X$_i = $ X$_{\textrm{min}}$.  If it is not, then X$_i$ is refered to as a local minimum.  Classically, evolutionary (or genetic) algorithms \cite{jong,forrest,srinivas,parsons} are one example strategy for overcoming local minima.  Similarly, quantum algorithms have also demonstrated success in this area for both annealing \cite{finnila,koshka} and gate-based \cite{wierichs,rivera,sack}.  
	
	The strategy for verifying a local versus global minimum using amplitude amplification can be seen by comparing the region $0.0029 \leq p_{\textrm{s}} \leq 0.00291$ in figures \ref{Fig.12} and \ref{Fig.13}.  For the linear QUBO corresponding to these figures, there exists a solution C(X$_i$) $= -1497$ which becomes maximally probable at $p_{\textrm{s}} \approx 0.002914$, followed by the next lowest solution C(X$_i$) $= -1491$ at $p_{\textrm{s}} \approx 0.002892$.  Because there are no binary combinations X$_i$ that can produce values $ -1492 \geq$ C(X$_i$) $ \geq -1496$, the $p_{\textrm{s}}$ region that $\textit{would}$ correspond to their solutions instead produces nothing measurably significant.  This can be seen by the low cumulative probabilities in figure \ref{Fig.12}, as well as experimentally in figure \ref{Fig.13} as a gap in red data points for this $p_{\textrm{s}}$ region across all three simulations.

\begin{figure}[h]            
		\centering
		\includegraphics[scale=.4]{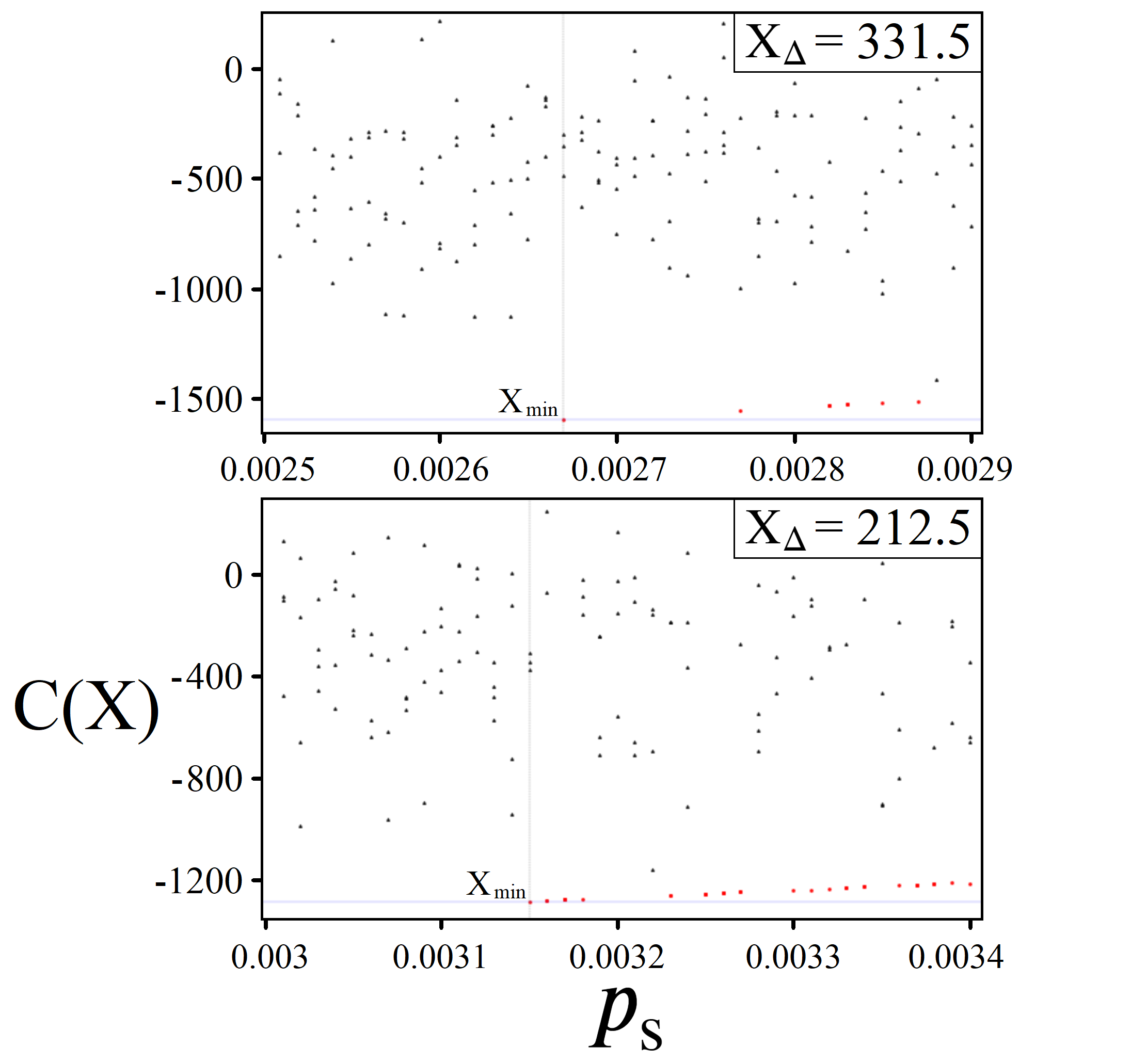}
		\caption{ Simulated measurement results for $p_{\textrm{s}}$ regions above and below the optimal point for finding $| \textrm{X}_{\textrm{min}} \rangle$.  Each plot corresponds to a different linear QUBO of size $N=25$, $k=4000$, with X$_{\Delta}$ values reported for each (top plot corresponds to the QUBO from figures \ref{Fig.9} - \ref{Fig.13}).  The point where X$_{\textrm{min}}$ is measured is indicated in both plots by the intersection of the blue (horizontal) and grey (vertical) lines.  Red-circle data points represent measurement results within the best $30$ minimizing solutions to C(X), otherwise as black triangles.  }
		\label{Fig.14}
	\end{figure}

	The ability for quantum to determine if an X$_i$ solution is locally or globally minimum is achieved by searching past the $p_{\textrm{s}}$ value corresponding to the solution.  Doing so will result in one of two outcomes: either a lower C(X$_j$) value will be probabilistically found (confirming X$_i$ was a local minimum), or the experimenter will only find random measurement results (X$_i$ was the global minimum).  Examples of this can be seen in figure \ref{Fig.14}, showcasing simulated measurement results as an experimenter searches past the optimal $p_{\textrm{s}}$ value for $| \textrm{X}_{\textrm{min}} \rangle$.

	The simulated experiments shown in figure \ref{Fig.14} were chosen to highlight both favorable (bottom) and unfavorable (top) cases for quantum.  The commonality between both experiments is that there is a clear point in $p_{\textrm{s}}$ (grey line) in which decreasing $p_{\textrm{s}}$ further results in only noisy random measurements.  However, determining this cutoff point using measurement results alone is challenging.  The top plot corresponds to the QUBO from figures \ref{Fig.10} - \ref{Fig.12}, which is the non-ideal situation in which there are significant gaps in solutions between the best $20$ minimizing C(X$_i$).  Experimentally this manifests as numerous $p_{\textrm{s}}$ regions that could be wrongly interpreted as the X$_{\textrm{min}}$ cutoff point.  Conversely, the bottom plot represents the ideal case where the best minimizing C(X$_i$) solutions are all closely clustered together.  This leads to a much more consistent correlation of measurement results leading to X$_{\textrm{min}}$, followed by an evident switch to randomness.
	
	The significance here is that amplitude amplification has an experimentally verifiable means for identifying the global minimum X$_{\textrm{min}}$ of a cost function.  Similarly, the same methodology can be in principle used to check for the existence of an X$_i$ solution corresponding to any given cost function value, which we discuss further in section VII.C.  However, the obvious drawback is that this verification technique relies on numerous amplitude amplification measurements finding nothing, which costs further runtime as well as being probabilistic.  As we discuss in the next section, a more realistic application of this quantum feature is to help steer a classical algorithm past local minima, leaving the veification of X$_{\textrm{min}}$ as joint effort between both quantum and classical.

	\section{Hybrid Solving} 
	
	The results of section V. were all features of amplitude amplification using $U_{\textrm{c}}$ that were found through classical simulations of quantum systems.  They represent the primary motivation of this study, which is to demonstrate amplitude amplifaction's potential and the conditions for which it can be experimentally realized.  By contrast, the discussions of section VI. here are more speculative.  Given all of the results from sections III. - V., we now discuss how the strengths and weaknesses of amplitude amplification synergize with a parallel classical computer.
	
	The plots shown in figures \ref{Fig.13} and \ref{Fig.14} represent a very non-optimal approach to finding X$_{\textrm{min}}$, functionally a quantum version of an exhaustive search.  If the ultimate goal is to solve a cost function problem as quickly as possible, then it is in our best interest to use any and all tools available.  This means using a quantum computer when it is advantageous, and similarly also recognizing when the use of a classical computer is more appropriate.  In this section we discuss this interplay between quantum and classical, and the situations in which an experimenter may favor one or the other.  Shown below in figure \ref{Fig.15} is the general outline of a variational amplitude amplification model which relies solely on quantum to produce X$_{\textrm{min}}$.
	
	\begin{figure}[h]            
		\centering
		\includegraphics[scale=.42]{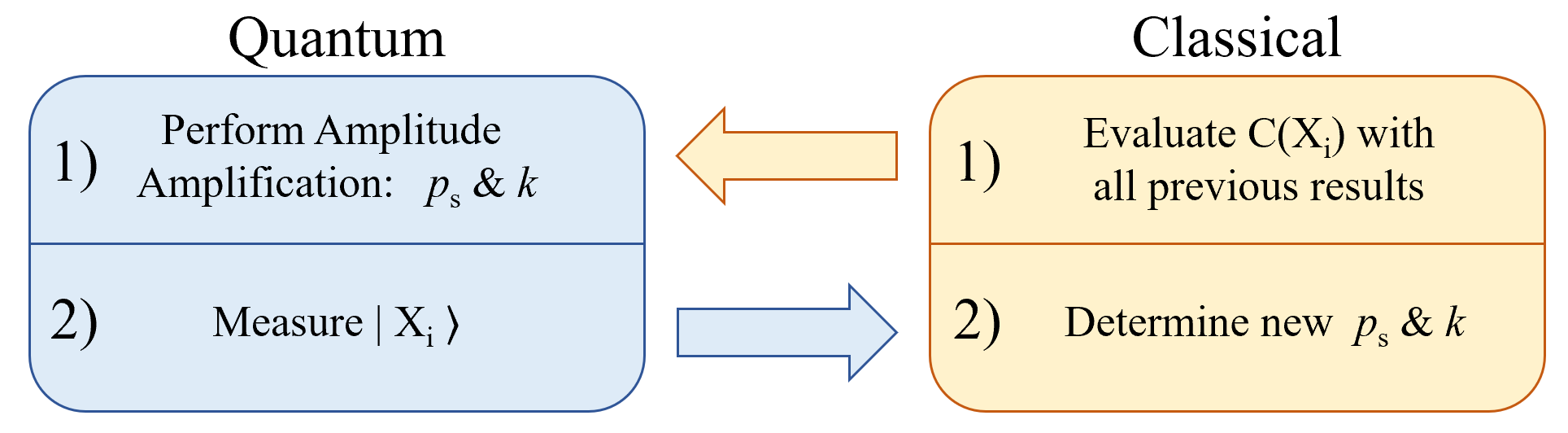}
		\caption{ The general outline of a variational amplitude amplification workflow.  Information from amplitude amplification in the form of measurements is fed to a classical optimizer between runs.  The optimizer then processes this information to supply the quantum computer with the next set of values $p_{\textrm{s}}$ and $k$, repeating this process until X$_{\textrm{min}}$ or another suitable solution is found. }
		\label{Fig.15}
	\end{figure}
	
	Given the current state of qubit technologies \cite{eisert,willsch,noiri}, performing one complete amplitude amplification circuit should be considered a scarce resource.  As such, it is the role of a classical optimizer to determine the most effective use of this resource, choosing $p_{\textrm{s}}$ and $k$ values which will probabilistically get the most value out of each attempt.  Determining optimal values to adjust a quantum circuit is the typical hybrid strategy found among other popular variational models of quantum computing \cite{qaoa,qaoa2,vqe}.  The majority of the computational workload is placed on the QPU (quantum processing unit), while a classical optimizer is used in between runs to adjust quantum circuit parameters accordingly.  As evidenced by figures \ref{Fig.13} and \ref{Fig.14}, this model is possible for amplitude amplification as well.  However, there is a different model of hybrid computing which  better utilizes both quantum and classical's strengths, shown below in figures \ref{Fig.16} and \ref{Fig.17}.

	\begin{figure}[h]            
		\centering
		\includegraphics[scale=.38]{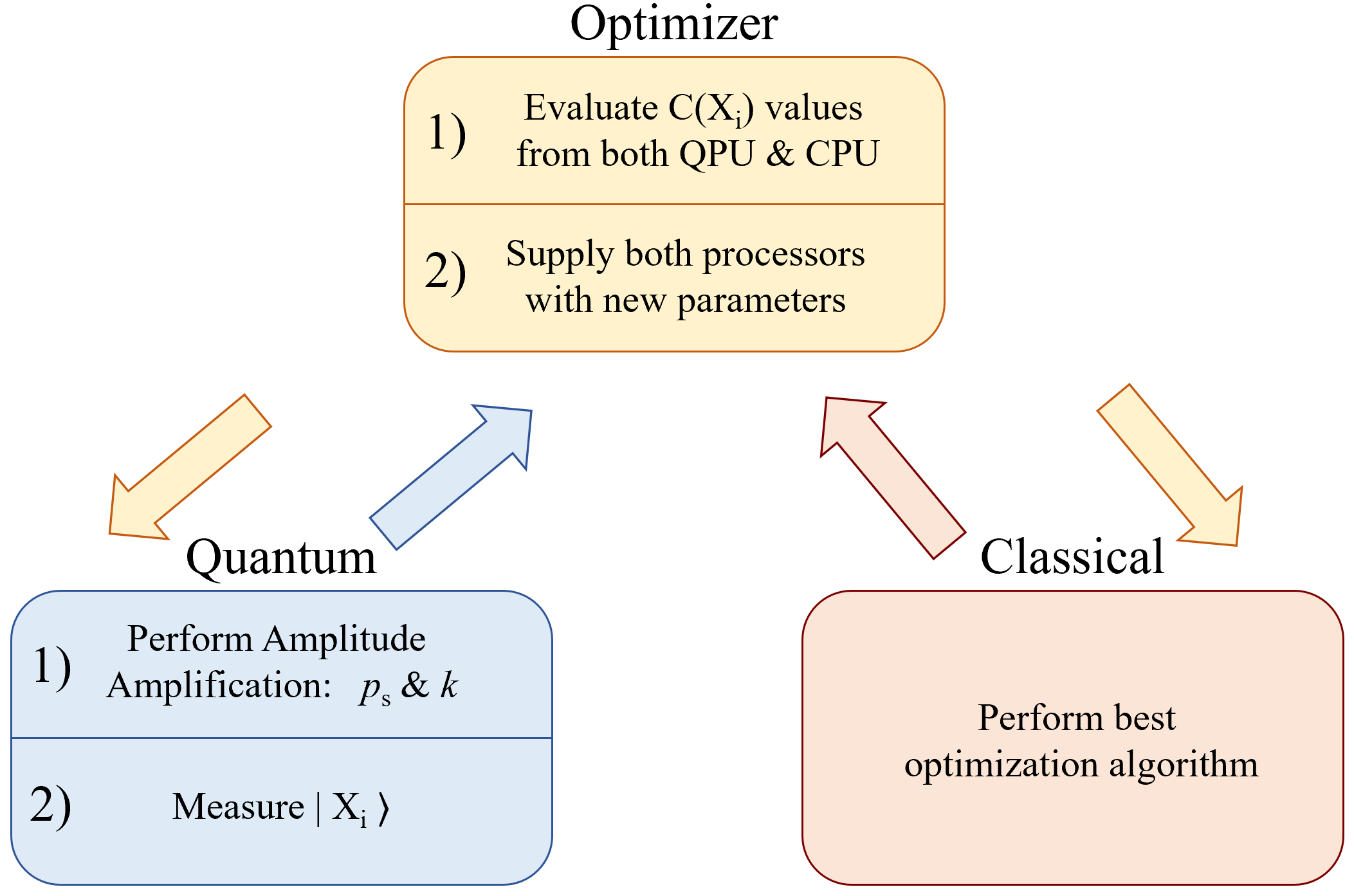}
		\caption{ Workflow of a hybrid model of computing, utilizing both a quantum and classical computer.  Both the QPU and CPU are run in parallel, and the information obtained from both are fed into the same classical optimizer, which in turn determines the most effective use for each processor. }
		\label{Fig.16}
	\end{figure}
	
	The advantage to hybrid computing using the model shown in figure \ref{Fig.16} is that both processors are working in tandem to solve the same problem, utilizing information gained from one another.  Information obtained through amplitude amplification measurements can be used to speedup a classical algorithm, and vice versa.  As we discuss further in the next subsection, this pairing of quantum and classical is maximally advantageous when the strengths of both computers compliment each other's weaknesses.

	\subsection{Supporting Greedy Algorithms}%
	
	One notable strength of classical computing is `greedy' algorithms, which oftentimes provide heuristic solutions for use cases ranging from biology and chemistry \cite{durand,zhang} to finance \cite{lin}.  Greedy algorithms are particularly viable for problems that possess certain structures which can be exploited \cite{korte}.  The key feature to these algorithms is that they focus on making locally optimal decisions which yield the maximal gain towards being optimal.  Consequently, they are very good at finding near optimal solutions quickly, but are also  prone to getting bottlenecked into local minima \cite{bang2}.
	
	The motivation for pairing amplitude amplification with a classical greedy algorithm is best exemplified by figures \ref{Fig.12} and \ref{Fig.13}.  The quantum states illustrated in these figures represent $|\textrm{X}_i \rangle$ states which rank as the $30^{\textrm{th}} - 80^{\textrm{th}}$ best minimizing solutions to C(X).  Under the right conditions it is reasonable to expect that a quantum computer could yield a solution in this range within $1-5$ amplitude amplification attempts.  The question then becomes how quickly a classical greedy algorithm could achieve the same feat?  Without problem specific structures to exploit, and as problem sizes scale like $2^N$, it becomes increasingly unlikely that classical can compete heuristically with quantum, which we argue is quantum's first advantage over classical in a hybrid model.
	
	Now, supposing that amplitude amplification does yield a low C(X$_i$) solution faster than classical, the problem then flips back to being classically advantageous.  This is because the X$_i$ solution provided by quantum is now new information available to the classical greedy algorithm.  As such, beginning the greedy approach from this new binary string is likely to yield even lower C(X$_i$) solutions in a time frame faster than amplitude amplification.  For example, this is the exact scenario in which genetic algorithms shine \cite{jong,forrest,srinivas,parsons,lin}, where a near-optimal solution is provided from which they can manipulate and produce more solutions.  And if a fast heuristic solution is all that is needed, then quantum's job is done, and the best minimal solution found by the classical greedy algorithm completes the hybrid computation.
	
	But if a heuristic solution is not enough, then we can continue to use a hybrid quantum-classical strategy for finding X$_{\textrm{min}}$. Referring back now to figures \ref{Fig.13} and \ref{Fig.14}, the strategy for quantum is to use multiple amplitude amplification trials and measurements to approximate the underlying correlation from figures \ref{Fig.10} and \ref{Fig.11}.  The fastest means for achieving this is to work in a $p_{\textrm{s}}$ region analogous to figure \ref{Fig.12}, where experimentally one has the highest probabilities of measuring useful information.  Simultaneously, the classical greedy algorithm can also find X$_i$ solutions in this area as it searches for X$_\textrm{min}$.  Knowledge of these X$_i$ solutions can be directly fed back to quantum, which can be used to predict $p_{\textrm{s}}$ values where solutions are known to exist, speeding up the process of determining a $p_{\textrm{s}}$ vs. C(X) correlation.  Thus, after quantum initially aided classical, subsequent information obtained from classical is then used to speed up quantum.
	
\begin{figure}[h]            
		\centering
		\includegraphics[scale=.43]{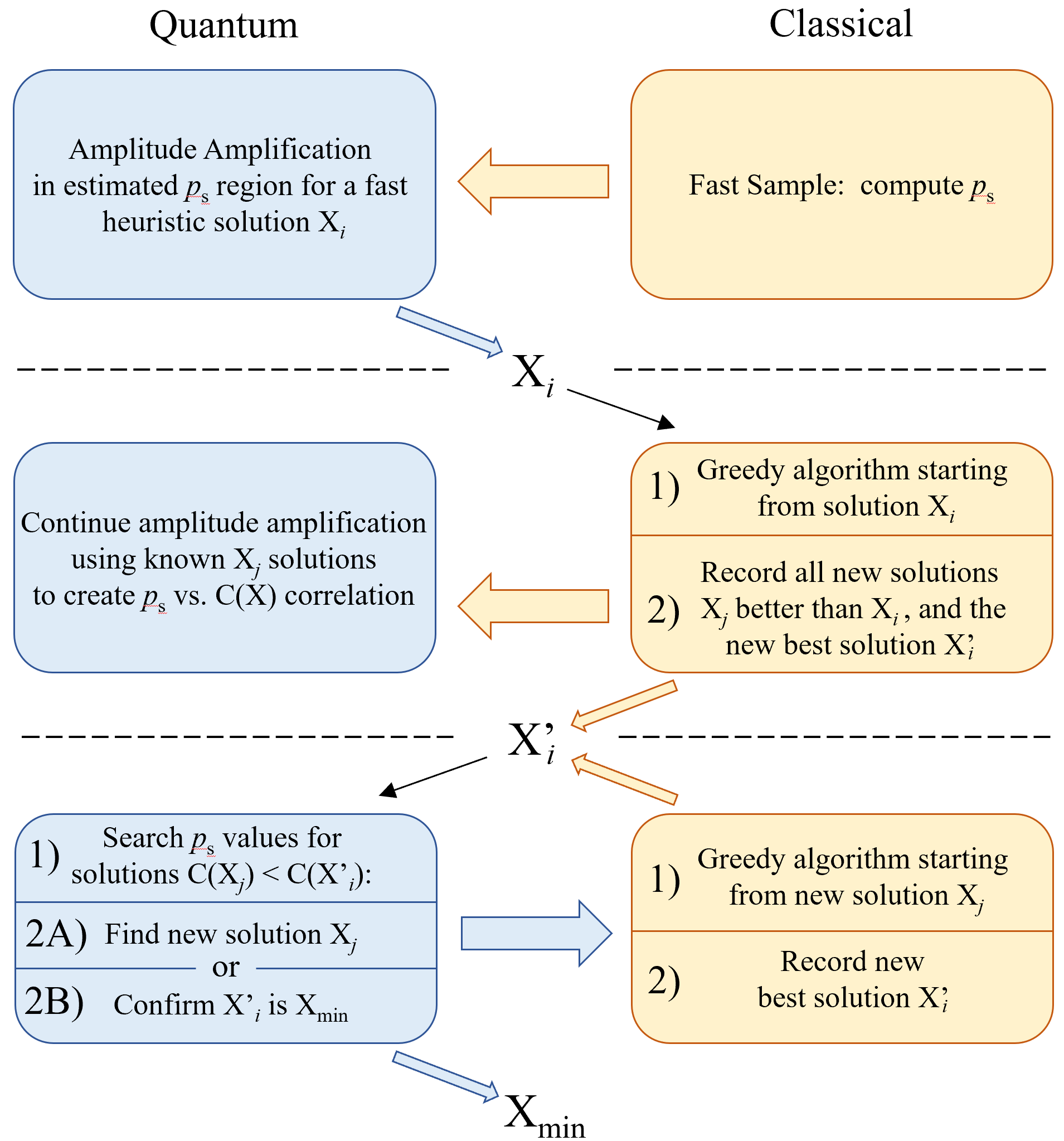}
		\caption{ Workflow for a hybrid model of computing between quantum amplitude amplification and a classical greedy algorithm.  The full strategy is broken up into three phases: 1) Amplitude amplification provides the first heuristic solution X$_i$. 2) A classical greedy algorithm uses X$_i$ to find a more optimal solution X'$_i$.  Simultaneously, other near optimal solutions X$_j$ are used to assist amplitude amplification in determining a $p_{\textrm{s}}$ vs. C(X) correlation (see figures \ref{Fig.10} - \ref{Fig.13}). 3) The correlation best-fit is used to predict $p_{\textrm{s}}$ values where solutions C(X$_j$) $<$ C(X'$_i$) must exist (or C(X$_j$) $>$ C(X'$_i$) for maximization problems).  Amplitude amplification attempts for these $p_{\textrm{s}}$ values will either produce a new best X$_j$ for the greedy classical algorithm to use, or confirm X'$_{i}$ $=$ X$_{\textrm{min}}$.  }
		\label{Fig.17}
	\end{figure}

	In the time it takes for quantum to experimentally verify enough $p_{\textrm{s}}$ and C(X$_i$) values to create a predictive correlation, we expect the classical algorithm to find a new lowest C(X$_i$) solution, labeled X'$_i$ in figure \ref{Fig.17}.  After investing additional trials into the amplitude amplification side of the computation, it is now time for quantum's second advantage: verifying local versus global minima.  Using an approximate $p_{\textrm{s}}$ vs C(X) best-fit, the quantum computer can skip directly to the $p_{\textrm{s}}$ value corresponding to best currently known X'$_i$ solution.  As discussed in section V.D, searching past this $p_{\textrm{s}}$ value will result in one of two outcomes.  Either the quantum computer will find a new best solution C(X$_j$) $<$ C(X'$_i$), or confirm that X'$_i$ is indeed the global minimum X$_{\textrm{min}}$.  In the former case, the greedy algorithm now starts again from the new lowest solution X$_j$, repeating this cycle between quantum and classical until X$_{\textrm{min}}$ is found. Figure \ref{Fig.17} below shows a workflow outline of this hyrbid strategy.
	
	The biggest advantage to using a hybrid model like shown in figure \ref{Fig.17} is that it can be adapted to each problem's uniqueness.  Problems with known fast heuristic techniques can lean on the classical side of the computation more heavily \cite{glover2,festa}, while classically difficult problems can put more reliance on quantum \cite{karp,garey}.  But above all else, this model of computation incorporates and synergizes the best known classical algorithms with quantum, rather than competing against them.

	\section{More Oracle Problems}

	All of the results from sections III. - V. were derived from linear QUBOs according to equations \ref{Eqn.1} - \ref{Eqn.4}.  However, these results can be applied to more challenging and realistic optimization problems provided that 1) all possible solutions can be encoded via phases by an appropriate oracle operation $U_{\textrm{c}}$, and 2) the distribution of all possible answers is suitable for boosting the solution we seek (gaussian, polynomial, exponential, etc. \cite{bench}).  Here we will briefly note some additional optimization problems which meet both of these criteria.

	\subsection{Weighted \& Unweighted Max-Cut}%
	
	The Maximum Cut problem (`Max-Cut') is famously NP-Hard \cite{karp}, where the objective is to partition every vertex in a graph $\mathbb{S}$ into two subsets $\mathbb{P}_1$ and $\mathbb{P}_2$ such that the number of edges between them is maximized.  In the weighted Max-Cut version of the problem, each edge is given a weight $w_{ij}$, and the goal is to maximize the sum of weights contained on edges between $\mathbb{P}_1$ and $\mathbb{P}_2$.  The unweighted Max-Cut problem has already been demonstrated as a viable use for amplitude amplification \cite{satoh}, which we will build upon further here via the weighted version.  Equation \ref{Eqn.31} below is the cost function C(X) for the weighted Max-Cut problem, which can be converted to the unweighted case by setting every edge weight $w_{ij} = 1$.  The binary variables $x_i$ here represent being partitioned into  $\mathbb{P}_1$ or $\mathbb{P}_2$.
	
	\begin{eqnarray}            
		\textrm{C}(\textrm{X}) &=&  \sum_{ \{ i , j \} \in \mathbb{S}}  w_{ij}  \hspace{0.02cm} | x_i - x_j |  \label{Eqn.31}  
	\end{eqnarray}
	
	\begin{figure}[h]            
		\centering
		\includegraphics[scale=.4]{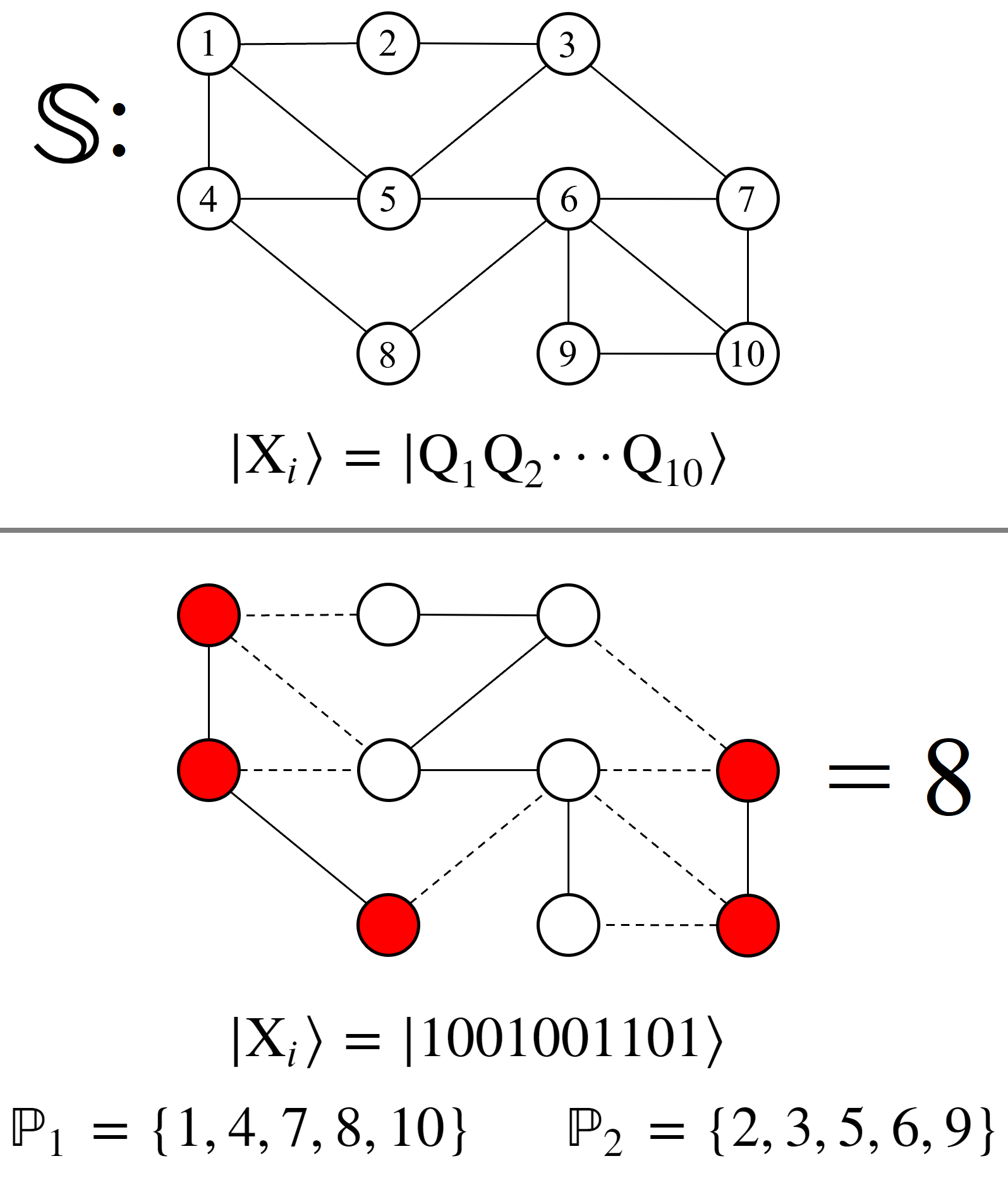}
		\caption{ (top) A graph $\mathbb{S}$ composed of $10$ nodes and $15$ connections.  Each node is labeled $1$ - $10$, corresponding to the qubits Q$_1$ - Q$_{10}$ shown below. (bottom) An example Max-Cut solution X$_i$, along with its quantum state representation $| \textrm{X}_i \rangle$. Nodes colored red correspond to the partition $\mathbb{P}_1$, quantum state $|1\rangle$, while nodes colored white correspond to partition $\mathbb{P}_2$, quantum state $|0\rangle$.  `Cuts' are represented in the graph as dashed lines, totaling $8$ for this example.}
		\label{Fig.18}
	\end{figure}
	
	Shown in figure \ref{Fig.18} is an example graph $\mathbb{S}$ and one of its solutions.  This example graph is composed of $10$ vertices, labeled $1$ - $10$, and a total of $15$ connecting edges.  Encoding this graph requires one qubit per vertex, where the basis states $| 1 \rangle$ and $| 0 \rangle$ represent belonging to the subsets $\mathbb{P}_1$ and $\mathbb{P}_2$ respectively.  See the bottom graph in figure \ref{Fig.18} for an example solution state.
	
	The cost oracle $U_{\textrm{c}}$ for solving Max-Cut must correctly evaluate all $2^N$ solution states $| \textrm{X}_i \rangle$ based on the edges of $\mathbb{S}$ according to equation \ref{Eqn.31}.  For example, if vertices $1$ and $2$ are partitioned into different sets, then $U_{\textrm{c}}$ needs to affect their combined states $|\textrm{Q}_1 \textrm{Q}_2 \rangle = |01\rangle$ and $| 10 \rangle$ with the correct phase, weighted or unweighted.  Just like figure \ref{Fig.3} from earlier, we can achieve this with a control-phase gate CP($\theta$),with the intent of scaling by $p_{\textrm{s}}$ later (see figure \ref{Fig.4}).  The caveat here is that we need this phase on $|01\rangle$ and $| 10 \rangle$, not $|11\rangle$, which means that additional X gates are required for the contruction of $U_{\textrm{c}}$, shown below in equation \ref{Eqn.32}.
	
	\begin{eqnarray}            
		\textrm{X}&=& \begin{bmatrix} 0 & 1 \\ 1 & 0 \end{bmatrix} \label{Eqn.32} 
	\end{eqnarray}
	
	For the complete $U_{\textrm{c}}$ quantum circuit which encodes the graph $\mathbb{S}$ in figure \ref{Fig.18}, please see appendix $\textbf{C}$. Once properly scaled by $p_{\textrm{s}}$, the solutions which are capable of boosting are determined by the underlying solution space distribution of the problem, which can be seen in figure \ref{Fig.19} below.  The histogram in this figure shows all $2^{10}$ C(X$_i$) solutions to the graph $\mathbb{S}$ from figure \ref{Fig.18}.  Even for a $10$ qubit problem size such as this, it is clear that the underlying solution space distribution shows gaussian-like structure.

	\begin{figure}[h]            
		\centering
		\includegraphics[scale=.4]{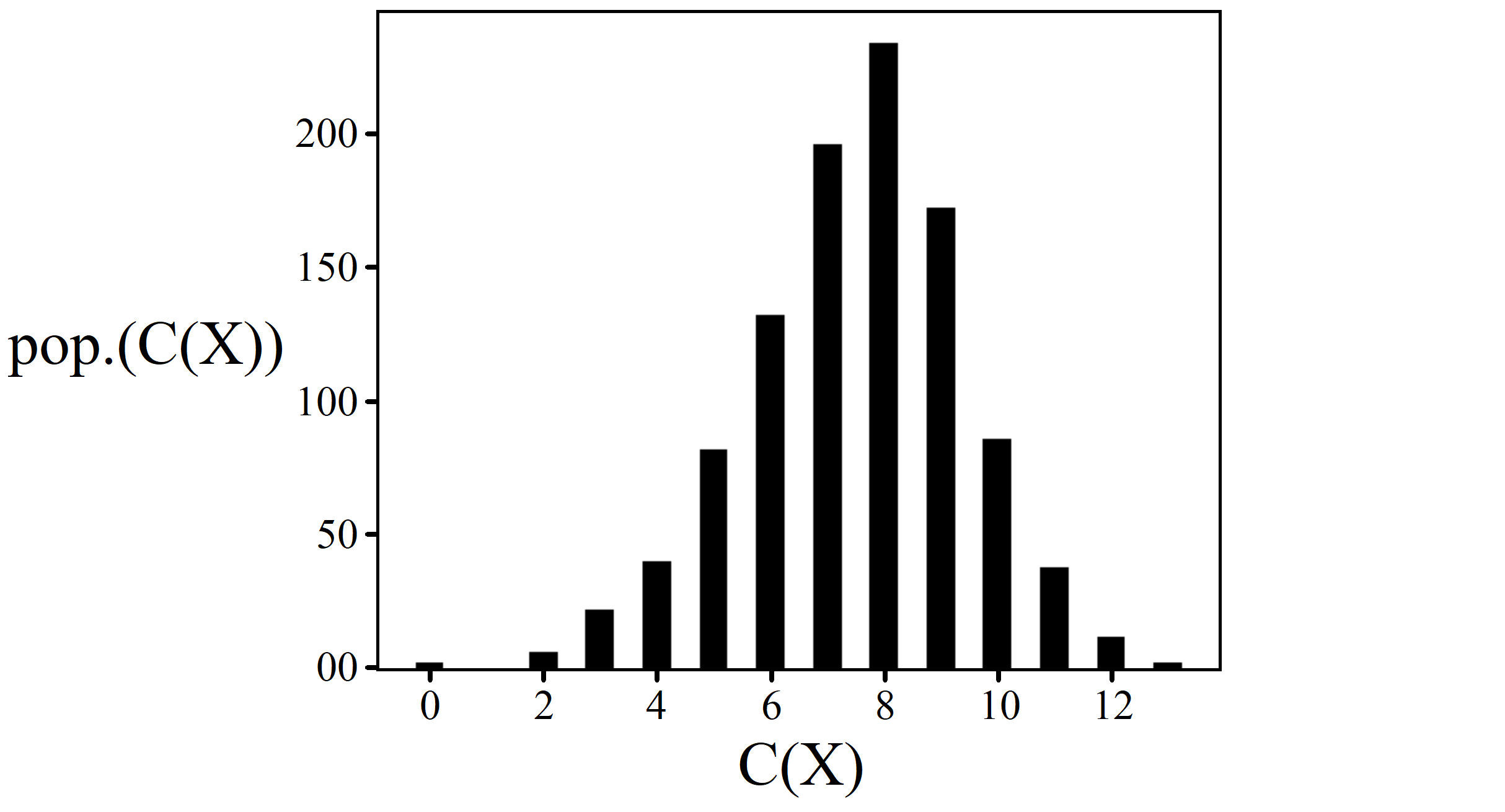}
		\caption{ A histogram of all $2^{10}$ solutions for an unweighted Max-Cut on graph $\mathbb{S}$ from figure \ref{Fig.18}.}
		\label{Fig.19}
	\end{figure}

	One interesting feature of Max-Cut is that all solutions come in equal and opposite pairs.  For example, the optimal solutions to $\mathbb{S}$ from figure \ref{Fig.19} are $|  0 1 0 0 1 0 1 1 10 \rangle$ and $|  1011010001 \rangle$, which both yield $13$ `cuts'.  Mathematically there is no difference between swapping all vertices in $\mathbb{P}_1$ and $\mathbb{P}_2$, but physically it means that there are always two optimal solution states.  Consequently, these states will always share the effect of amplitude amplification together, which is something an experimenter must be aware of when choosing iterations $k$.
	
	Finally, moving from the unweighted to weighted version of Max-Cut increases the problem's difficulty, but notably does not change the circuit depth of $U_{\textrm{c}}$.  Rather than using $\theta = 1 $ for all of the control-phase gates, each $\theta$ now corresponds to a weighted edge $w_{ij}$ of the graph.  Similar to the QUBO distributions shown in figure \ref{Fig.7}, this increase in complexity allows for more distinct C(X$_{i}$) solutions, and consequently more variance in features such as $\sigma '$ and X$_{\Delta}$.

	\subsection{Graph Coloring}%
	
	A similar optimization problem to Max-Cut is Graph Coloring, also known as Vertex Coloring \cite{karp}, which extends the number of allowed partition sets $\mathbb{P}_i$ up to any integer number $k$ ($k=2$ is equivalent to Max-Cut).  Given a graph of vertices and edges $\mathbb{S}$, the goal is to assign every vertex to a set $\mathbb{P}_i$ such that the number of edges between vertices within the same sets is minimized.  Shown below in equation \ref{Eqn.33} is the cost function C(X) for a $k$-coloring problem, where the values of each vertex $x_i$ are no longer binary, but can take on $k$ different integer values.  The quantity inside the parentheses is equal to $1$ if $x_i = x_j$, and $0$ for all other combinations $x_i \neq x_j$. Just like with Max-Cut, setting all $w_{ij} = 1$ is the unweighted version of the problem.

	\begin{eqnarray}            
		\textrm{C}(\textrm{X}) &=&  \sum_{ \{ i , j \} \in \mathbb{S}}  w_{ij} \left( 1 -  \hspace{0.02cm} \Big{\lceil}  \frac{ | x_i - x_j | }{k}   \Big{\rceil} \right) \label{Eqn.33}  
	\end{eqnarray}
	
	The name `coloring' is in reference to the problem's origins, whereby the sets $\mathbb{P}_{i}$ all represent different colors to be applied to a diagram, such as a map.  Shown below in figure \ref{Fig.20} is an example picture composed of overlapping shapes, where each section must be assigned one of $k$ colors such that the number of adjacent sections with the same color is minimized.  Example solutions for $k=3$ and $k=4$ are shown, along with their vertex and quantum state representations of the problem.
	
	\begin{figure}[h]            
		\centering
		\includegraphics[scale=.4]{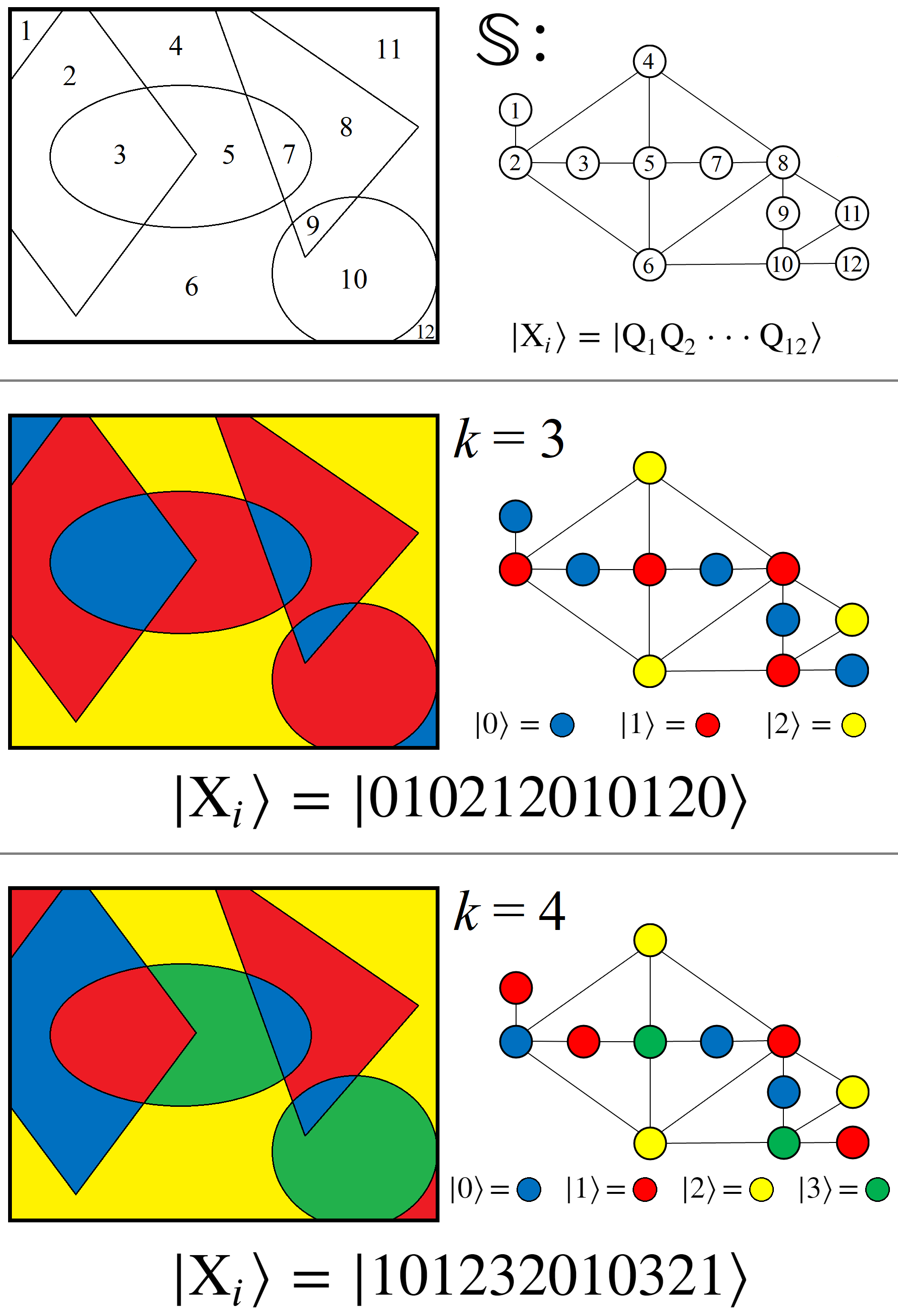}
		\caption{ (top) On the left, a two dimensional bounded picture with overlapping geometric shapes. On the right, a graph $\mathbb{S}$ representing the $12$ distinct regions of the picture as nodes.  Connections between nodes in $\mathbb{S}$ represent regions in the picture which share a border, not counting adjacent corners. (middle) A $k=3$ coloring example, with a corresponding $d=3$ qudit state representation.  (bottom) A $k=4$ coloring example, with a corresponding $d=4$ qudit state representation.   }
		\label{Fig.20}
	\end{figure}
	
	In order to encode graph coloring as an oracle $U_{\textrm{c}}$, the choice of $k$ determines whether qubits or another form of quantum computational unit is appropriate.  While qubits are capable of producing superposition between two quantum states, qudits are the generalized unit of quantum information capable of achieving superposition between $d$ states \cite{wang,lanyon,luo,niu}.  To see why this is necessary, let us compare the $k=3$ and $4$ examples from figure \ref{Fig.20}, and the quantum states needed to represent partitioning each vertex.  
	
	For $k=4$, we need four distinct quantum states to represent a vertex belonging to one of the $\mathbb{P}_i$ partitions.  While a single qubit can't do this, a pair of qubits can.  Thus, every vertex in $\mathbb{S}$ can be encoded as a pair of qubits, letting the basis states $| 00 \rangle$, $| 01 \rangle$, $| 10 \rangle$, and $| 11 \rangle$ each represent a different color.  Alternatively, we could use a $d=4$ qudit to represent each vertex, assigning each partition a unique basis state: $| 0 \rangle$, $| 1 \rangle$, $| 2 \rangle$, or $| 3 \rangle$, such as the state shown in figure \ref{Fig.20}.  Mathematically the two encodings are identical, so the choice between whether to use qubits or qudits is a matter of experimental realization (i.e. which technology is easier to implement).
	
	For $k=3$ however, two qubits is too many states, and a single qubit is not enough.  So in order to represent three colors exactly in quantum, the appropriate unit is a `qutrit' (the common name for a $d=3$ qudit).  Similarly, all prime numbers $d$ can only be encoded as their respective $d$-qudit, while all composite values can be built up from combinations of smaller qudits.  Once an appropriate mixed-qudit quantum system is determined, constructing $U_{\textrm{c}}$ is the same as the Max-Cut problem from earlier, but now with $k$ state-state interactions. For an example of qudit quantum circuits and their use for amplitude amplification, please see Wang et al. \cite{wang} and our previous work on the Traveling Salesman problem\cite{koch2}.

	\subsection{Subset Sum}%
	
	For all of the oracles discussed thus far, the circuit depth and total gate count for $U_{\textrm{c}}$ is determined by the size and connection complexity of $\mathbb{S}$, the graphical representation of the problem.  By contrast, the simplest possible quantum circuit that can be used as $U_{\textrm{c}}$ corresponds to the Subset Sum problem \cite{karp}.  The cost function for this problem is given in equation \ref{Eqn.34}.
	
	\begin{eqnarray}            
		\textrm{C}(\textrm{X}) &=&  \sum_{ i }^N   W_{i}  \hspace{0.02cm} x_i   \label{Eqn.34}  
	\end{eqnarray}
	
	Rather than optimizing equation \ref{Eqn.34}, which is trivial, the Subset Sum problem is to determine if there exists a particular combination such that C(X$_i$) = $T$, where $T$ is some target sum value.  The boolean variables $x_i$ represent which $W_i$ values to use as contributors to the sum. Figure \ref{Fig.21} below shows an $N=10$ example.  Note that this problem is equally applicable to any of the other oracles discussed thus far, whereby we can ask if a target value $T$ exists for some graph $\mathbb{S}$.

	\begin{figure}[h]            
		\centering
		\includegraphics[scale=.46]{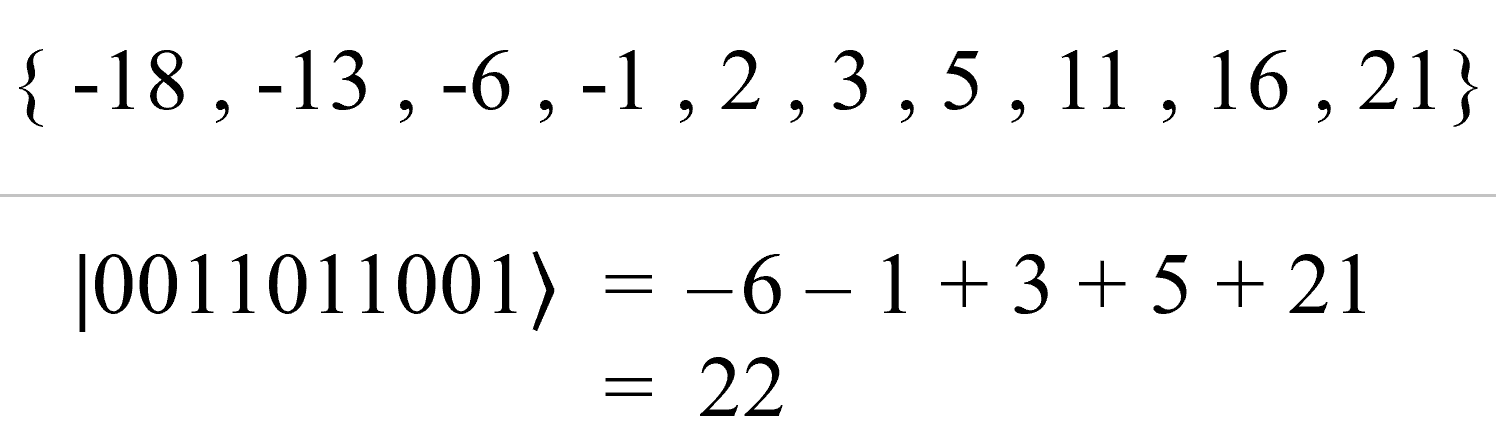}
		\caption{ (top) A set of $10$ integer values, shown in ascending order, from which we are intereted in solving the Subset Sum problem for $T=22$. (bottom) An example solution state $| \textrm{X}_i \rangle$ corresponding to the cost function value C(X) = $22$. }
		\label{Fig.21}
	\end{figure}

	The reason why equation \ref{Eqn.34} is the simplest $U_{\textrm{c}}$ oracle one can construct is because the cost function doesn't contain any weights $w_{ij}$ that depend on two variables.  Consequently, the construction of $U_{\textrm{c}}$ doesn't use any $2$-qubit phase gates CP($\theta$), instead only requiring a single qubit phase gate P($\theta$) for every qubit.  In principle, all of these single qubit operations can be applied in parallel, such as in figure \ref{Fig.3}, which means that the circuit depth of $U_{\textrm{c}}$ is exactly one.
	
	Although this is the most gate efficient $U_{\textrm{c}}$, using it to solve the Subset Sum problem comes with some limitations.  Firstly, it can only solve for $T$ values within a limited range.  This is illustrated by the results of figure \ref{Fig.11}, which demonstrate that amplitude amplification can only produce meaningful probabilities of measurement up to a certain threshold away from X$_{\textrm{min}}$ or X$_{\textrm{max}}$.  Consequently, one can only use $U_{\textrm{c}}$ here if the target sum value $T$ is within this threshold distance from the extrema.
	
	The second limitation to consider is the discussion from section V.D, whereby the information of whether a state C(X$_i$) $= T$ exists or not may rely on measurements finding nothing.  Previously we discussed how an experimenter might iteratively decrease $p_{\textrm{s}}$ and eventually expect to find regions where cost function values do not exist (see figure \ref{Fig.14}) as one approaches X$_{\textrm{min}}$.  Here things are easier, since an experimenter can test for $p_{\textrm{s}}$ values above and below where C(X$_i$) $= T$ (except for the case where $T$ is the global extrema).  Using a $p_{\textrm{s}}$ vs. C(X) correlation in this manner can confirm exactly where the $p_{\textrm{s}}$ value for C(X$_i$) $= T$ must be.  Testing this $p_{\textrm{s}}$ window will then either confirm the existence of a solution for $T$ via a measurement, or conversely confirm no solution exists through multiple trials of random measurement results.

	\section{Conclusion}
	
	The results of this study demonstrate that the gate-based model of amplitude amplification is a viable means for solving combinatorial optimization problems, particularly QUBOs.  The ability to encode information via phases and let the $2^N$ superposition of qubits naturally produce all possible combinations is a feature entirely unique to quantum.  Harnessing this ability into a useful algorithmic form was the primary motivation for this study, and as we've shown, is not without its own set of challenges.  In particular, the discussions of sections IV.A \& IV.B highlight that this algorithm is not a `one size fits all' strategy that can be blindly applied to any QUBO.  Depending on how the numerical values of a given problem form a solution space distribution, it may simply be impossible for amplitude amplification to find one extrema or the other.  Figure \ref{Fig.8} shows that at least one of the extrema solutions is always viable for quantum to find, it just may not happen to be the one that is of interest to the experimenter.
	
	For cases where the desired solution is well-suited for quantum to find, that is $| \textrm{X}_{\textrm{min}} \rangle$ or $| \textrm{X}_{\textrm{max}} \rangle$ is capable of achieving a high probability of measurement, a different challenge lies in finding the correct $p_{\textrm{s}}$ value to use in order to boost these states.  However, the results of section V. illustrate that this challenge is solvable via quantum measurement results.  If the best an experimenter could do is simply guess at $p_{\textrm{s}}$ and hope for success, then amplitude amplification would not be a practical algorithm.  But the correlations shown in figures \ref{Fig.10} and \ref{Fig.11} illustrate that that is not the case, and that information about $p_{\textrm{s}}$ can be experimentally learned and used to find extrema solutions.  How quickly this information can be experimentally produced, analyzed, and used is exactly how quickly quantum can find the optimal solution, which is an open question for further research.
	
	While the free parameter $p_{\textrm{s}}$ can be considered the bottleneck of our algorithm for finding optimal solutions, there is a second important metric by which we can judge the usefulness of amplitude amplification: as a heuristic algorithm.  A major finding of this study is depicted in figure \ref{Fig.12}, which shows that there is a wide range of $p_{\textrm{s}}$ values for which quantum can find an answer within the best $1 - 5$\% of all solutions.  And as we demonstrated in section IV.C with sampling, it is not unrealistic that a classical computation can estimate this $p_{\textrm{s}}$ region very quickly.  The question then becomes how does this compare to classical greedy algorithms, and how quickly can they achieve the same feat in a timescale compared to quantum's O($\frac{\pi}{4} \sqrt{N/M}$) for problem sizes of $2^N$.  The answer to this question will vary from problem to problem, but certainly in some cases such as highly interconnected QUBOs we view this as the first practical use for amplitude amplification.
	
	And finally, there is one important sentiment from section VI that we would like to reiterate again here, namely that amplitude amplification is a technique that benefits tremendously from working in parallel with a classical computer.  The information learned through quantum measurements can equally be of use to speeding up quantum as well as a classical algorithm.  And vice versa, information learned through a classical greedy algorithm can be used to speed up quantum.  The goal of this hyrbid computing  model is to utilize the advantages both computers have to offer, and ultimately to find optimal solutions faster than either computer can achieve alone.  Understanding which optimization problems this scenario may be applicable to is the future direction of our research.

	\section*{Acknowledgments}
	
	Any opinions, findings, conclusions or recommendations expressed in this material are those of the author(s) and do not necessarily reflect the views of AFRL.
	
	\section*{Data \& Code Availability}
	
	The data and code files that support the findings of this study are available from the corresponding author upon reasonable request.

	\clearpage
	
	
	\clearpage

	\appendix

	\section{QUBO data}
	
	For this study, linear QUBOs as defined in equation \ref{Eqn.4} were created using a uniform random number generator for node and edge weights according to equations \ref{Eqn.2} and \ref{Eqn.3}.  The total number of QUBOs produced and analyzed to create figure \ref{Fig.6} is given below in table \ref{Tab.2}.  Every QUBO was simulated through amplitude amplification, and the $p_{\textrm{s}}$ value which yielded the highest probability of measurement for $| \textrm{X}_{\textrm{min}} \rangle$ was recorded.
	
	\begin{table}[h]
		\begin{tabular}{|c|c|}
			\hline
			$\hspace{0.35cm} N \hspace{0.35cm}$  & $\hspace{0.2cm}$ \# of QUBOs studied $\hspace{0.2cm}$  \\ \hline
			17   &   5000  \\ \hline
			18   &   3000  \\ \hline
			19   &   2000  \\ \hline
			20   &   1500  \\ \hline
			21   &   1200  \\ \hline
			22   &   1000  \\ \hline
			23   &   1000  \\ \hline
			24   &   600  \\ \hline
			25   &   500  \\ \hline
			26   &   400  \\ \hline
			27   &   100  \\ \hline
		\end{tabular}
		\caption{ Table of values showing the number of linear QUBOs generated and studied per size $N$. }
		\label{Tab.2}
	\end{table}

	\section{Linear Regression}
	
	In order to determine how linearly correlated the data points in figure \ref{Fig.10} were, a regression best-fit was performed according to equations \ref{Eqn.B1} - \ref{Eqn.B5} below.  The collection of (x,y) data points D in equation \ref{Eqn.B1} corresponds to the ($p_{\textrm{s}}$,C(X$_i$)) points in the figure.  The resulting linear correlation factor R is reported at the top of figure \ref{Fig.10}.
	
	\begin{eqnarray}           
		\textrm{D} &=& ( (x_1,y_1) , (x_2 , y_2) , ... , (x_N , y_N ) ) \label{Eqn.B1} \\ 
		\bar{X} &=& \sum_i^N x_i   \hspace{0.6cm}   \bar{X^2} =  \sum_i^N (x_i)^2  \\  
		\bar{Y} &=&  \sum_i^N y_i  \hspace{0.6cm}   \bar{Y^2} =  \sum_i^N (y_i)^2 \\ 
		\bar{XY} &=&  \sum_i^N x_i \cdot y_i \\
		R &=& \frac{N \bar{XY} - \bar{X} \bar{Y}}{\sqrt{ (N \bar{X^2} - (\bar{X})^2)(N \bar{Y^2} - (\bar{Y})^2) }} \label{Eqn.B5} 
	\end{eqnarray}
	
	$ \hspace{1cm} $
	
	\section{Max-Cut Circuit}
	
\begin{figure}[h]            
		\centering
		\includegraphics[scale=.6]{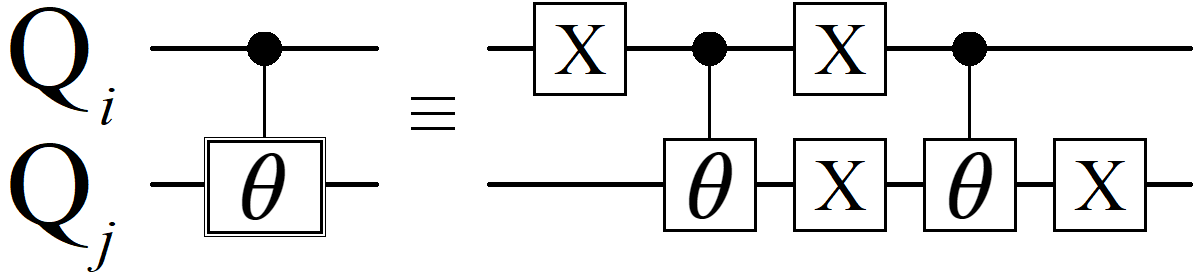}
		\caption{ Quantum circuit which achieves the $2$-qubit unitary from equation \ref{Eqn.C1}.}
		\label{Fig.22}
	\end{figure}

\begin{figure}[h]            
		\centering
		\includegraphics[scale=.46]{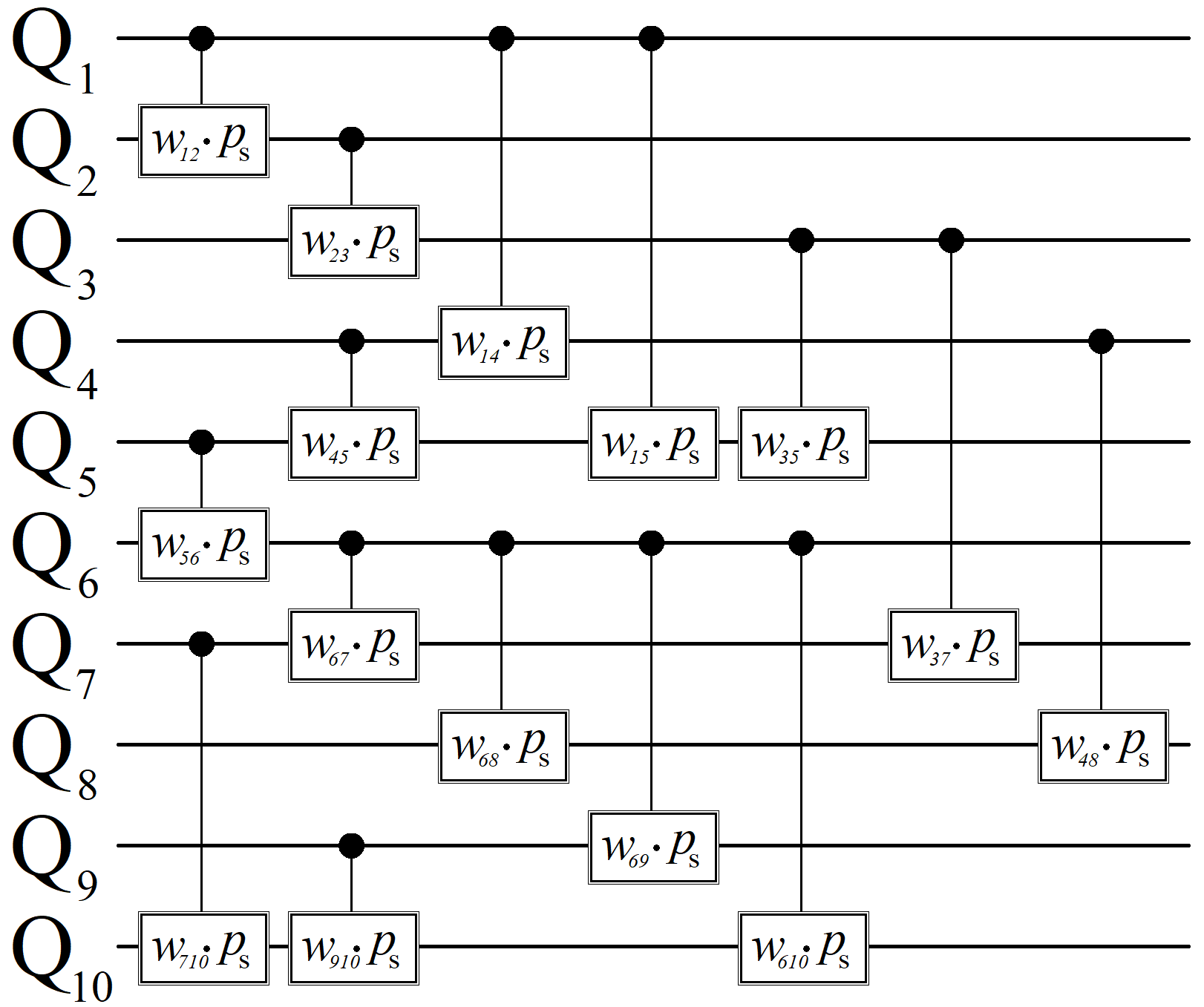}
		\caption{ Quantum circuit which achieves the oracle $U_{\textrm{c}}$ corresponding to $\mathbb{S}$ from figure \ref{Fig.18}, for the Max-Cut problem. Each gate shown represents one of the $15$ connections in $\mathbb{S}$, corresponding to the custom gate defined in figure \ref{Fig.22}.  The placement of gates shown here are spread out for clarity, while a real implementation could be more parallelized.}
		\label{Fig.23}
	\end{figure}

	To illustrate how any graph structure $\mathbb{S}$ can be encoded as an oracle $U_{\textrm{c}}$, figure \ref{Fig.23} below is the quantum circuit corresponding to $\mathbb{S}$ from figure \ref{Fig.18}.  Because this oracle needs to represent a Max-Cut problem (weighted or unweighted), the states which must acquire phases are $|01\rangle$ and $|10\rangle$.  To make the circuit less cluttered, let us define the custom gate given in figure \ref{Fig.22}.

	The quantum circuit shown in figure \ref{Fig.22}, drawn similar to a CP($\theta$) gate but with an extra box around it, is an operation which achieves the following unitary:
	
	\begin{eqnarray}           
		U ( &&\alpha |00 \rangle + \beta |01 \rangle + \gamma |10 \rangle + \rho |11 \rangle \hspace{0.3cm})  \label{Eqn.C1} \\
		=&&\alpha |00 \rangle + e^{i \theta} \beta |01 \rangle + e^{i \theta} \gamma |10 \rangle + \rho |11 \rangle \nonumber 
	\end{eqnarray}
	
	The unitary $U$ from equation \ref{Eqn.C1} is the required operation for representing the cost oracle given in equation \ref{Eqn.31}.  If two nodes (qubits) share a connection in $\mathbb{S}$, then a `cut' corresponds to them being partitioned into different sets, which is represented by the qubit states $|0\rangle$ and $|1\rangle$.  Figure \ref{Fig.23} uses the operation in figure \ref{Fig.22} to create the complete $U_{\textrm{c}}$ circuit for encoding all $15$ connections in $\mathbb{S}$.
	
\begin{thebibliography}{99}
		\bibitem{grover} L. K. Grover, A fast quantum mechanical algorithm for database search. arXiv: 9605043 (1996)
		\bibitem{boyer} M. Boyer, G. Brassard, P. Hoyer, A. Tapp, Tight bounds on quantum searching. Fortschritte der Physik \textbf{46}, p.493-506 (1998)
		\bibitem{bennett} C. H. Bennett, E. Bernstein, G. Brassard, U. Vazirani, Strengths and weaknesses of quantum computing. SIAM Journal on Computing \textbf{26} (5), p.1510-1523 (1997)
		\bibitem{farhi} E. Farhi and S. Gutmann, Analog analogue of a digital quantum computation. Phys. Rev. A \textrm{57}, 2403 (1998)
		\bibitem{brassard1} G. Brassard, P. Hoyer, A. Tapp, Quantum counting. 25th Intl. Colloquium on Automata, Languages, and Programming (ICALP), LNCS 1443, p. 820-831, (1998)
		\bibitem{brassard2} G. Brassard, P. Hoyer, M. Mosca, A. Tapp, Quantum amplitude amplification and estimation. Quantum Computation and Quantum Information: AMS Contemporary Mathematics \textbf{305}, p.53-74 (2002)
		\bibitem{childs} A. M. Childs, J. Goldstone, Spatial search by quantum walk. Phys. Rev. A \textbf{70}, 022314 (2004)
		\bibitem{ambainis} A. Ambainis, Variable time amplitude amplification and a faster quantum algorithm for solving systems of linear equations. arXiv: 1010.4458 (2010)
		\bibitem{singleton} R. L. Singleton Jr., M. L. Rogers, D. L. Ostby, Grover's algorithm with diffusion and amplitude steering. arXiv: 2110.11163 (2021)
		\bibitem{koch2} D. Koch, M. Cutugno, S. Karlson, S. Patel, L. Wessing, P. M. Alsing, Gaussian amplitude amplification for quantum pathfinding. Entropy \textbf{24} (7), 963 (2022)
		\bibitem{lloyd} S. Lloyd, Quantum search without entanglement. Phys. Rev. A \textbf{61}, 010301(R) (1999)
		\bibitem{viamontes} G. F. Viamontes, I. L. Markov, J. P. Hayes, Is quantum search practical? arXiv: 0405001 (2004)
		\bibitem{reg} O. Regev and L. Schiff, Impossibility of a quantum speed-up with a faulty oracle. arXiv: 1202.1027
		\bibitem{seidel} R. Seidel, C. K-U. Becker, S. Bock, N. Tcholtchev, I-D. Gheorge-Pop, M. Hauswirth, Automatic generation of grover quantum oracles for arbitrary data structures. Quantum Sci. Tech. \textbf{8}, 025003 (2023) 
		\bibitem{nielsen} M. A. Nielsen, I. L. Chuang, \textit{Quantum Computation and Quantum Information}, Cambridge University Press, p.249 (2000)
		\bibitem{bang} J. Bang, S. Yoo, J. Lim, J. Ryu, C. Lee, J. Lee, Quantum heuristic algorithm for traveling salesman problem. J. Korean Phys. Soc. \textbf{61}, 1944 (2012) 
		\bibitem{satoh} T. Satoh, Y. Ohkura, R. V. Meter, Subdivided phase oracle for NISQ search algorithms. IEEE Transactions on Quantum Engineering (2020) 
		\bibitem{bench} N. Benchasattabuse, T. Satoh, M. Hajdušek, R. V. Meter, Amplitude amplification for optimization via subdivided phase oracle. arXiv: 2205.00602 (2022)
		\bibitem{shyamsundar} P. Shyamsundar, Non-boolean quantum amplitude amplification and quantum mean estimation. arXiv: 2102.04975 (2021)
		\bibitem{gilliam} A. Gilliam, S. Woerner, C. Gonciulea, Grover adaptive search for constrained polynomial binary optimization. Quantum \textbf{5}, 428 (2021)
		\bibitem{roy} T. Roy, L. Jiang, D. I. Schuster, Deterministic Grover search with a restricted oracle. Phys. Rev. Research \textbf{4}, L022013 (2022)
		\bibitem{plek} K. Plekhanov, M. Rosenkranz, M. Fiorentini, M. Lubasch, Variational quantum amplitude estimation. Quantum \textbf{6}, 670 (2022)
		\bibitem{long1} G. L. Long, W. L. Zhang, Y. S. Li, L. Niu, Arbitrary phase rotation of the marked state cannot be used for Grover's quantum search algorithm. Commun. Theor. Phys. \textbf{32} (3), p.335 (1999)
		\bibitem{long2} G. L. Long, Y. S. Li, W. L. Zhang, L. Niu, Phase matching in quantum searching. Phys. Lett. A \textbf{262}, p.27-34 (1999)
		\bibitem{hoyer} P. Hoyer, Arbitrary phases in quantum amplitude amplification. Phys. Rev. A \textbf{62}, 052304 (2000)
		\bibitem{younes} A. Younes, Towards more reliable fixed phase quantum search algorithm. Applied Math. \& Info. Sciences \textbf{1} (7), 10 (2013)
		\bibitem{li1} T. Li, W-S. Bao, W-Q. Lin, H. Zhang, X-Q. Fu, Quantum search algorithm based on multi-phase. Chinese Phys. Lett. \textbf{31} (5), 050301 (2014)
		\bibitem{guo} Y. Guo, W. Shi, Y. Wang, J. Hu, Q-learning-based adjustable fixed-phase quantum Grover search algorithm. Journal of the Physical Society of Japan \textbf{86} (2), 024006 (2017)
		\bibitem{song} P. H. Song and I. Kim, Computational leakage: Grover's algorithm with imperfections. European Phys. Jour. D \textbf{23} (2000)
		\bibitem{pomeransky} A. A. Pomeransky, O. V. Zhirov, D. L. Shepelyansky, Phase diagram for the Grover algorithm with static imperfections. European Phys. Jour. D \textbf{31} (2004)
		\bibitem{janmark} J. Janmark, D. A. Meyer, T. G. Wong, Global symmetry is unnecessary for fast quantum search. Phys. Rev. Lett. \textbf{112}, 210502 (2014)
		\bibitem{kochenberger} G. Kochenberger, J-K. Hao, F. Glover, M. Lewis, Z. Lu, H. Wang, Y. Wang, The unconstrained binary quadratic programming problem: a survey. Journal of Combinatorial Optimization \textbf{28} (1), p.58–81 (2014)
		\bibitem{lucas} A. Lucas, Ising formulations of many NP problems. Front. Phys. \textbf{12}, 2 (2014)
		\bibitem{glover} F. Glover, G. Kochenberger, Y. Du, A tutorial on formulating and using QUBO models. arXiv: 1811.11538 (2018)
		\bibitem{date} P. Date, D. Arthur, L. Pusey-Nazzaro, QUBO formulations for training machine learning models. Scientific Reports \textbf{11} (1), 10029 (2021)
		\bibitem{herman} D. Herman, C. Googin, X. Liu, A. Galda, I. Safro, Y. Sun, M. Pistoia, Y. Alexeev, A survey of quantum computing for finance. arXiv: 2201.02773 (2022)
		\bibitem{date2} P. Date, R. Patton, C. Schuman, T. Potok, Efficiently embedding QUBO problems on adiabatic quantum computers. Quantum Inf. Process. \textrm{18} (4), 117 (2019)
		\bibitem{ushijima} H. Ushijima-Mwesigwa, C. F. A. Negre, S. M. Mniszewski, Graph partitioning using quantum annealing on the D-Wave system. arXiv: 1705.03082 (2017)
		\bibitem{pastorello} D. Pastorello, E. Blanzieri, Quantum annealing learning search for solving QUBO problems. Quantum Inf. Process. \textbf{18}, 10 (2019)
		\bibitem{cruz} W. Cruz-Santos, S. E. Venegas-Andraca, M. Lanzagorta, A QUBO formulation of minimum multicut problem instances in trees for D-Wave quantum annealers. Scientific Reports \textbf{9} (1), 17216 (2019)
		\bibitem{qaoa} E. Farhi, J. Goldstone, S. Gutmann, A quantum approximate optimization algorithm. arXiv:1411.4028 (2014)
		\bibitem{qaoa2} S. Hadfield, Z. Wang, B. O'Gorman, E. G. Rieffel, D. Venturelli, R. Biswas, From the quantum approximate optimization algorithm to a quantum alternating operator ansatz. Algorithms \textbf{12} (2), 34 (2019)
		\bibitem{guerreschi} G. G. Guerreschi, A. Y. Matsuura, QAOA for max-cut requires hundreds of qubits for quantum speed-up. Scientific Reports \textbf{9}, 6903 (2019)
		\bibitem{guerreschi2} G. G. Guerreschi, Solving quadratic unconstrained binary optimization with divide-and-conquer and quantum algorithms. arXiv: 2101.07813 (2021)
		\bibitem{streif} M. Streif, M. Leib, Comparison of QAOA with quantum and simulated annealing. arXiv: 1901.01903 (2019)
		\bibitem{gabor} T. Gabor, M. L. Rosenfeld, S. Feld, C. Linnhoff-Popien, How to approximate any objective function via quadratic unconstrained binary optimization. arXiv: 2204.11035 (2022)
		\bibitem{pelofske} E. Pelofske, A. Bartschi, S. Eidenbenz, Quantum annealing vs. QAOA: 127 qubit higher-order ising problems on NISQ computers. arXiv: 2301.00520 (2023)
		\bibitem{bernoulli} J. Bernoulli, \textit{Ars Conjectandi}, Basileae: Thurnisiorum. (1713)
		\bibitem{laplace} P. S. Laplace, Mémoire sur les approximations des formules qui sont fonctions de très grands nombres et sur leur application aux probabilités. Mémoires de l'Académie Royale des Sciences de Paris, \textbf{10} (1810)
		\bibitem{gauss} C. F. Gauss, \textit{Theoria Motus Corporum Coelestium in Sectionibus Conicis Solem Ambientium}, Hamburg: Friedrich Perthes and I.H. Besser (1809)
		\bibitem{vqe} A. Peruzzo, J. McClean, P. Shadbolt, M-H. Yung, X-Q. Zhou, P. J. Love, A. Aspuru-Guzik, J. L. O'Brien, A variational eigenvalue solver on a quantum processor. Nature Communications \textbf{5}, 4213 (2014)
		\bibitem{durand} K. Nieman, H. Durand, S. Patel, D. Koch, and P. M. Alsing, Application of quantum computing amplitude amplification techniques for solving problems in control and optimization. journal pending. (2023)
		\bibitem{jong} K. D. Jong, Learning with genetic algorithms: an overview. Machine Language \textbf{3}, p.121-139 (1988)
		\bibitem{forrest} S. Forrest, Genetic algorithms: principles of natural selection applied to computation. Science \textbf{261}, 5123 (1993)
		\bibitem{srinivas} M. Srinivas, L. M. Patnaik, Genetic algorithms: a survey. IEEE Computer \textbf{27}, p.17-26 (1994)
		\bibitem{parsons} R. J. Parsons, S. Forrest, C. Burks, Genetic algorithms, operators, and DNA fragment assembly. Machine Learning \textbf{21}, 11-33 (1995)
		\bibitem{finnila} A. B. Finnila, M. A. Gomez, C. Sebenik, C. Stenson, J. D. Doll, Quantum annealing: a new method for minimizing multidimensional functions. Chemical Physics Letters \textbf{219}, p.343-348 (1994)
		\bibitem{koshka} Y. Koshka, M. A. Novotny, Comparison of D-Wave quantum annealing and classical simulated annealing for local minima determination. IEEE Journal on Selected Areas in Information Theory \textbf{1}, 2 (2020)
		\bibitem{wierichs} D. Wierichs, C. Gogolin, M. Kastoryano, Avoiding local minima in variational quantum eigensolvers with the natural gradient optimizer. Phys. Rev. Research \textbf{2}, 043246 (2020)
		\bibitem{rivera} J. Rivera-Dean, P. Huembeli, A. Acin, J. Bowles, Avoiding local minima in variational quantum algorithms with neural networks. arXiv: 2104.02955 (2021)
		\bibitem{sack} S. H. Sack, M. Serbyn, Quantum annealing initialization of the quantum approximate optimization algorithm. Quantum \textbf{5}, 491 (2021)
		\bibitem{eisert} J. Eisert, D. Hangleiter, N. Walk, I. Roth, D. Markham, R. Parekh, U. Chabaud, E. Kashefi, Quantum certification and benchmarking. Nature Reviews Physics \textbf{2}, p.382-390 (2020)
		\bibitem{willsch} D. Willsch, M. Willsch, C. D. G. Calaza, F. Jin, H. De Raedt, M. Svensson, K. Michielsen, Benchmarking advantage and D-Wave 2000Q quantum annealers with exact cover problems. Quantum Inf. Process. \textbf{21}, 141 (2022)
		\bibitem{noiri} A. Noiri, K. Takeda, T. Nakajima, T. Kobayashi, A. Sammak, G. Scappucci, S. Tarucha, Fast universal quantum gate above the fault-tolerance threshold in silicon. Nature \textbf{601}, 7893 p.338-342 (2022)
		\bibitem{zhang} Z. Zhang, S. Schwartz, L. Wagner, W. Miller, A greedy algorithm for aligning DNA sequences. Journal of Comp. Biology \textbf{7}, p.203-214 (2004)
		\bibitem{lin} L. Lin, L. Cao, J. Wang, C. Zhang, The applications of genetic algorithms in stock market data mining optimisation. WIT Trans. on Info. and Comm. Tech. \textbf{33} (2004)
		\bibitem{korte} B. Korte, L. Lovasz, Mathematical structures underlying greedy algorithms. Fundamentals of Comp. Theory (1981)
		\bibitem{bang2} J. Bang-Jensen, G. Gutin, A. Yeo, When the greedy algorithm fails. Discrete Optimization \textbf{1} (2), p.121-127 (2004)
		\bibitem{glover2} F. Glover, G. Gutin, A. Yeo, A. Zverovich, Construction heuristics for the asymmetric TSP. European Journ. of Operational Research \textbf{129}, 3 (2001)
		\bibitem{festa} P. Festa, P. M. Pardalos, M. G. C. Resende, C. C. Ribeiro, Randomized heuristics for the Max-Cut problem. Optimization Methods and Software \textbf{17}, 6 (2002)
		\bibitem{karp} R. Karp, Reducibility among combinatorial problems. \textit{Proceedings of a symposium on the complexity of computer computations}, Yorktown Heights, New York (1972)
		\bibitem{garey} M. R. Garey, D. S. Johnson, L. Stockmeyer, Some simplified NP-complete graph problems. Theoretical Computer Science \textrm{1}, 3 p.237-267 (1976)
		\bibitem{wang} Y. Wang, Z. Hu, B. C. Sanders, S. Kais, Qudits and high-dimensional quantum computing.  Front. Phys. \textbf{10}, 8 (2020)
		\bibitem{lanyon} B. P. Lanyon, M. Barbieri, M. P. Almeida, T. Jennewein, T. C. Ralph, K. J. Resch, G. J. Pryde, J. L. O'Brien, A. Gilchrist, A. G. White, Quantum computing using shortcuts through higher dimensions. Nature Physics \textbf{5}, 134 (2009)
		\bibitem{luo} M-X. Luo and X-J. Wang, Universal quantum computation with qudits. Sci. China Phys. Mechanics \& Astronomy \textbf{57} (9), p.1712–1717 (2014)
		\bibitem{niu} M. Y. Niu, I. L. Chuang, J. H. Shapiro, Qudit-Basis Universal Quantum Computation Using $\chi^2$ Interactions. Phys. Rev. Lett. \textbf{120}, 160502 (2018)
		
		
		
		
		
		
		
		
		
		
	\end{thebibliography}
\end{document}